\documentclass[12pt]{article}
\pdfoutput=1
\usepackage{amsmath}
\usepackage{color}
\def\a{\alpha}
\def\b{\beta}

\def\d{\delta}
\def\e{\epsilon}

\def\g{\gamma}
\def\p{\psi}

\def\r{\rho}
\def\s{\sigma}

\def\be{\begin{equation}}
\def\ee{\end{equation}}
\def\arr{\begin{array}{rll}}
\def\ea{\end{array}}
\def\bea{\begin{eqnarray}}
\def\eea{\end{eqnarray}}
\def\N2{$N{=}2$}

\def\>{\rangle}
\def\<{\langle}
\def\+{\dagger}
\def\={\ =\ }

\begin{document}
%\large
\renewcommand{\thefootnote}{\fnsymbol{footnote}}
\begin{titlepage}
\setcounter{page}{0}

\begin{center}
{\LARGE\bf Remarks on $D(2,1;a)$ super-- }\\
\vskip 0.4cm
{\LARGE\bf Schwarzian derivative }\\
\vskip 1.5cm
$
\textrm{\Large Anton Galajinsky}^{a}  ~  \textrm{\Large  and Ivan Masterov}^{b}
$
\vskip 0.7cm
${}^{a}${\it
Tomsk Polytechnic University, 634050 Tomsk, Lenin Ave. 30, Russia} \\
\vskip 0.2cm
${}^{b}${\it
Tomsk State University of Control Systems and Radioelectronics, 634050 Tomsk, Russia} \\

\vskip 0.2cm
{e-mails:
galajin@tpu.ru, ivan.v.masterov@tusur.ru}
\vskip 0.5cm

\end{center}
\vskip 1cm
\begin{abstract} \noindent
It was recently demonstrated that $\mathcal{N}=1,2,3,4$ super--Schwarzian derivatives can be constructed by applying the method of nonlinear realisations to
finite--dimensional superconformal groups $OSp(1|2)$, $SU(1,1|1)$, $OSp(3|2)$, $SU(1,1|2)$, respectively, thus avoiding the use of superconformal field theory techniques.
In this work, a similar construction is applied to the exceptional supergroup $D(2,1;a)$, which describes the most general $\mathcal{N}=4$ supersymmetric extension of $SL(2,R)$, with the aim to study possible candidates for a $D(2,1;a)$ super--Schwarzian derivative. 

\end{abstract}


\end{titlepage}

\renewcommand{\thefootnote}{\arabic{footnote}}
\setcounter{footnote}0

\noindent
{\bf 1. Introduction}\\

\noindent
Over the past few decades, extensive studies of the $AdS/CFT$--correspondence brought into focus a plethora of interesting field theories, which hold invariant under conformal transformations. Most recent in the chain of dualities is the Schwarzian theory which connects to
two--dimensional models of gravity.\footnote{The literature on the subject is rather extensive.  For good recent accounts and further references see \cite{BBN,HITZ}.} In this context, the Schwarzian derivative and its supersymmetric counterparts play the central role.

The super--Schwarzian derivatives were introduced within the framework of $\mathcal{N}$--extended superconformal field theory in the mid--eighties \cite{F,Cohn,Sch,MU}. They showed up when computing the anomalous term in
a finite superconformal transformation of the super stress--energy tensor.
Mathematicians link them to central extensions of infinite--dimensional Lie superalgebras (see \cite{MD} and references therein), which also revealed the bound $\mathcal{N}\leq 4$ on an admissible number of supersymmetry parameters \cite{ChK}.

Given a superconformal diffeomorphism $t'=\rho(t,\theta)$, $\theta'_i=\psi_i (t,\theta)$ of ${\mathcal{R}}^{1|\mathcal{N}}$ superspace\footnote{A more precise definition of a superconformal diffeomorphism is given below in Sect. 2. Note also that,
because conformal transformations in ${\mathcal{R}}^{1|\mathcal{N}}$ involve the inversion $t \to \frac{1}{t}$,
$SL(2,R)$ does not act globally on ${\mathcal{R}}^{1|\mathcal{N}}$, but rather on ${\mathcal{S}}^{1|\mathcal{N}}$.}, which is parametrized by a real bosonic coordinate $t$ and real fermionic variables $\theta_i$, $i=1,\dots,\mathcal{N}$,
a super--Schwarzian derivative $S[\psi(t,\theta);t,\theta]$ acts upon the fermionic superfield $\psi_i(t,\theta)$ specifying
the transformation in the Grassmann--odd sector. In general, $S[\psi(t,\theta);t,\theta]$ is required to obey two defining properties. Firstly, given the change of the argument $\psi_i (t,\theta) \to \Omega_i (t',\theta')$,  a specific composition law should link $S[\Omega(t',\theta');t,\theta]$ to a combination of $S[\psi(t,\theta);t,\theta]$ and $S[\Omega(t',\theta');t',\theta']$. Secondly, solving the equation $S[\psi(t,\theta);t,\theta]=0$, one should reproduce a finite-dimensional superconformal transformation acting in the Grassmann--odd sector of ${\mathcal{R}}^{1|\mathcal{N}}$. In view of the composition law, the latter determines the symmetry group of an $\mathcal{N}$--extended super--Schwarzian derivative.

In a series of recent works \cite{AG1,AG,GK,AG2}, an alternative procedure of constructing $\mathcal{N}=0,1,2,3,4$ super--Schwarzian derivatives was elaborated, which consisted in applying the method of nonlinear realisations \cite{CWZ} to finite dimensional (super)conformal groups $SL(2,R)$, $OSp(1|2)$, $SU(1,1|1)$, $OSp(3|2)$, and $SU(1,1|2)$, respectively. The derivatives were built directly in terms of the Maurer--Cartan invariants, thus avoiding the use of conformal field theory techniques or analysis of cocycles of infinite--dimensional Lie superalgebras.

As is well known \cite{FSS}, the most general $\mathcal{N}=4$ supersymmetric extension of $SL(2,R)$ is given by the exceptional supergroup $D(2,1;a)$. Structure relations of the corresponding Lie superalgebra (see Appendix B) involve an arbitrary real parameter $a$. To the best of our knowledge, despite the fact that $D(2,1;a)$ admits an infinite--dimensional extension \cite{STVP}, no attempt was made to build a $D(2,1;a)$ super--Schwarzian derivative by applying superconformal field theory methods.

The goal of this work is to extend the recent group--theoretic analysis of the $SU(1,1|2)$ super--Schwarzian derivative in \cite{GK}, which corresponds to $a=-1$, to the case of arbitrary values of $a$. Surprisingly enough,
the construction of a $D(2,1;a)$ super--Schwarzian derivative obeying the two defining properties mentioned above turns out to be problematic. Yet, a reasonable alternative is proposed and its properties are established.

The paper is organised as follows.

In Sect. 2, a superconformal diffeomorphism  $t'=\rho(t,\theta,\bar\theta)$, $\theta'_\alpha=\psi_\alpha (t,\theta,\bar\theta)$, where $t$ is a real Grassmann--even variable and $\theta_\alpha$ are complex Grassmann--odd coordinates with $\alpha=1,2$, is considered. It is argued that the conventional chirality constraint on the fermionic superfield $\bar{\mathcal{D}}_{\alpha} \psi_\beta=0$, which normally follows from the condition that the covariant derivative transforms homogeneously under the superconformal isomorphisms, i.e. $\mathcal{D}^{\alpha} = (\mathcal{D}^{\alpha}\psi_{\beta}){\mathcal{D}'}^{\beta}$, turns out to be incompatible with $D(2,1;a)$ symmetry. A weaker condition is proposed, $\mathcal{D}^{\alpha} = (\mathcal{D}^{\alpha}\psi_{\beta}){\mathcal{D}'}^{\beta} + (\mathcal{D}^{\alpha}\bar{\psi}^{\beta}){\bar{\mathcal{D}}'}_{\beta}$, in which the covariant derivatives $\mathcal{D}'^{\beta}$ and $\bar{\mathcal{D}}'_{\beta}$ get mixed up. The requirement that the covariant derivatives algebra is preserved under the superconformal diffeomorphism
gives rise to extra quadratic constraints on $\psi_\alpha$ and $\bar\psi^\alpha$.

In Sect. 3, infinitesimal $D(2,1;a)$ transformations acting upon the form of the superfields $\rho$ and $\psi_\alpha$ are constructed by applying the method of nonlinear realisations. They are subsequently used for verifying the invariance of constraints imposed upon $\psi_\alpha$ as well as for analysing the infinitesimal limit of finite $D(2,1;a)$ transformations, which result from setting a generalised $D(2,1;a)$ super--Schwarzian derivative to zero.

Sect. 4 is devoted to the construction of $D(2,1;a)$ invariants along the lines in \cite{GK}. First, each generator in the corresponding Lie superalgebra is accompanied by a Goldstone superfield of the same Grassmann parity and a conventional group--theoretic element $\tilde g$ is introduced. Then the odd analogues of the Maurer--Cartan invariants ${\tilde g}^{-1} \mathcal{D}^\alpha \tilde g$ are computed. They are subsequently used to link some of the Goldstone superfields entering $\tilde g$ to a single fermionic superfield $\psi_\alpha$ as well as to construct a $D(2,1;a)$ analogue of the $\mathcal{N}=4$ super-Schwarzian derivative in \cite{MU,GK}. It is demonstrated that such a candidate lacks the conventional composition law unless extra nonlinear constraints are imposed upon $\psi_\alpha$.

In Sect. 5, a covariant projection method is used to solve the nonlinear constraints explicitly. A finite form of a $D(2,1;a)$ transformation acting in the Grassmann--odd sector of $\mathcal{R}^{1|4}$ is obtained. It is argued that a natural generalisation of the $\mathcal{N}=4$ super--Schwarzian derivative to the $D(2,1;a)$ case turns out to be trivial. As a matter of fact, it is superseded by one of the nonlinear constrains, which was imposed upon $\psi_\alpha$ when securing the composition law in Sect. 4.

An alternative definition of a $D(2,1;a)$ super--Schwarzian derivative and its peculiar features are discussed in Sect. 6. The composition law and the change under $D(2,1;a)$ transformations are established, which resemble the way in which the covariant derivative is transformed under the generalised superconformal diffeomorphism.
It is suggested that the difficulty in defining a $D(2,1;a)$ super--Schwarzian derivative with conventional properties is related to the fact that the chirality condition on the fermionic superfield $\psi_\alpha$ is incompatible with $D(2,1;a)$ symmetry.

In Sect. 7, we draw a parallel with the $\mathcal{N}=3$ case, which was recently studied in \cite{AG2} within a similar framework. Like for $D(2,1;a)$, the basic fermionic superfield does not obey the chirality condition.
A generalised $\mathcal{N}=3$ super--Schwarzian derivative is introduced and its properties are established. It is shown that the conventional $\mathcal{N}=3$ super--Schwarzian derivative \cite{Sch} can be constructed in terms of the generalised object. A new $OSp(3|2)$ invariant is proposed.

In the concluding Sect. 8, we summarise our results and discuss possible further developments.

Appendix A contains our spinor conventions.

Structure relations of the Lie superalgebra associated with the exceptional superconformal group $D(2,1;a)$ are gathered in Appendix B.

The $D(2,1;a)$ Maurer--Cartan invariants are exposed in Appendix C.

Throughout the paper, summation over repeated indices is understood.

\vspace{0.5cm}

\noindent
{\bf 2. Generalised superconformal diffeomorphisms}\\

\noindent
An $\mathcal{N}$--extended super--Schwarzian derivative \cite{F,Cohn,Sch,MU} is intimately connected with finite--dimensional superconformal transformations acting in $\mathcal{R}^{1|\mathcal{N}}$ superspace parametrized by $(t,\theta_i)$, $i=1,\dots,\mathcal{N}$. In order to obtain the latter from the former, one considers a generic
superdiffeomorphism
\be
t'=\rho(t,\theta), \qquad \theta'_i=\psi_i (t,\theta),
\ee
which is specified by a real bosonic superfield $\rho$ and a set of real fermionic superfields $\psi_i$, and then confines oneself to a subgroup of superconformal diffeomorphisms, under which the covariant derivative ${\mathcal{D}}_i$ transforms homogeneously \cite{F}
\be\label{hom}
{\mathcal{D}}_i=\left({\mathcal{D}}_i \psi_j\right) {\mathcal{D}}'_j.
\ee
Eq. (\ref{hom}) yields constraints on $\rho$ and $\psi_i$ (see the discussion in \cite{Sch}). In particular, $\rho$ is fixed provided $\psi_i$ is known.

Note that for even $\mathcal{N}$ one usually introduces complex Grassmann--odd coordinates $\theta_\alpha$, $\alpha=1,\dots,\frac{\mathcal{N}}{2}$, in which case (\ref{hom}) implies the chirality condition ${\bar{\mathcal{D}}}_\alpha \psi_\beta=0$.

Having solved the constraints, which follow from (\ref{hom}), one then identifies $\psi_i$ with the argument of an $\mathcal{N}$--extended super--Schwarzian derivative and sets the latter to vanish. The resulting fermionic superfield coincides with a finite--dimensional superconformal transformation acting in the Grassmann--odd sector of $\mathcal{R}^{1|\mathcal{N}}$ superspace, while $\rho$ is fixed from (\ref{hom}) provided $\psi_i$ is known.

Trying to extend such a consideration to the exceptional superconformal group $D(2,1;a)$, one immediately reveals a problem. The chirality condition
${\bar{\mathcal{D}}}_\alpha \psi_\beta=0$ and its complex conjugate partner $\mathcal{D}^{\alpha}\bar\psi^\beta=0$,
which would result from (\ref{hom}), turn out to be too restrictive. Indeed, $D(2,1;a)$ transformations involve $SU(2)$ subgroup which interchanges $\psi_\alpha$ and $\bar\psi_\alpha$ (see $\delta_{p_{\pm}}$ transformations in Eq. (\ref{tr}) below). Had we chosen the chirality condition to hold, $D(2,1;a)$ symmetry would entail $\mathcal{D}^{\alpha}\psi_\beta=0$, thus reducing $\psi_\beta$ to a constant.

A natural way out is to admit the weaker conditions
\bea\label{const}
&&
\mathcal{D}^{\alpha} = (\mathcal{D}^{\alpha}\psi_{\beta}){\mathcal{D}'}^{\beta} + (\mathcal{D}^{\alpha}\bar{\psi}^{\beta}){\bar{\mathcal{D}}'}_{\beta}, \qquad
\bar{\mathcal{D}}_{\alpha} = (\bar{\mathcal{D}}_{\alpha}\psi_{\beta}){\mathcal{D}'}^{\beta} + (\bar{\mathcal{D}}_{\alpha}\bar{\psi}^{\beta}){\bar{\mathcal{D}}'}_{\beta},
\eea
where $\mathcal{D}^{\alpha}$, $\bar{\mathcal{D}}_{\alpha}$ are the covariant derivatives in $\mathcal{R}^{1|4}$ superspace
\bea\label{cd}
&&
\mathcal{D}^{\alpha} = \frac{\vec\partial}{\partial\theta_{\alpha}} + i\bar{\theta}^{\alpha} \frac{\partial}{\partial t}, \qquad \bar{\mathcal{D}}_{\alpha} = \frac{\vec\partial}{\partial\bar{\theta}^{\alpha}} + i\theta_{\alpha}\frac{\partial}{\partial t},
\eea
which obey
\begin{align}\label{cda}
&
\{\mathcal{D}^{\alpha}, \mathcal{D}^{\beta}\}=0, &&   \{\bar{\mathcal{D}}_{\alpha},\bar{\mathcal{D}}_{\beta} \}=0, && \{\mathcal{D}^{\alpha},\bar{\mathcal{D}}_{\beta} \}=2 i {\delta_\beta}^\alpha \partial_t,
\\[2pt]
&
[{\mathcal{D}}^2,{\bar{\mathcal{D}}}_\alpha]=-4i {\mathcal{D}}_\alpha \partial_t, &&
[{\bar{\mathcal{D}}}^2,{\mathcal{D}}^\alpha]=-4i {\bar{\mathcal{D}}}^\alpha \partial_t, &&
[{\bar{\mathcal{D}}}^2,{\mathcal{D}}^2]=-4i\left({\mathcal{D}}^\alpha {\bar{\mathcal{D}}}_\alpha-{\bar{\mathcal{D}}}_\alpha {\mathcal{D}}^\alpha \right) \partial_t,
\nonumber
\end{align}
where $\partial_t=\frac{\partial}{\partial t}$.

Note that, after implementing the generalised superconformal diffeomorphism (\ref{const}), the covariant derivatives $\mathcal{D}'^{\beta}$ and $\bar{\mathcal{D}}'_{\beta}$ get mixed up. This fact will have an impact on the properties of a generalised $D(2,1;a)$ super--Schwarzian derivative to be introduced below.

From Eq. (\ref{const}) one gets the constraints
\bea\label{const1}
&&
\mathcal{D}^{\alpha}\rho = i\psi_{\beta}\mathcal{D}^{\alpha}\bar{\psi}^{\beta} + i\bar{\psi}^{\beta}\mathcal{D}^{\alpha}\psi_{\beta}, \qquad
\bar{\mathcal{D}}_{\alpha}\rho = i\psi_{\beta}\bar{\mathcal{D}}_{\alpha}\bar{\psi}^{\beta} + i\bar{\psi}^{\beta}\bar{\mathcal{D}}_{\alpha}\psi_{\beta},
\eea
which allow one to fix  $\rho$
\bea\label{const2}
&&
\partial_t\rho = -i\psi_{\alpha} \partial_t \bar{\psi}^{\alpha} - i\bar{\psi}^{\alpha} \partial_t \psi_{\alpha} + \frac{1}{2}\mathcal{D}\psi\bar{\mathcal{D}}\bar\psi + \frac{1}{2}\mathcal{D}\bar{\psi}\bar{\mathcal{D}}\psi,
\eea
provided $\psi_\alpha$ is known. Here and below we use the abbreviations
\bea\label{abbr}
&&
\mathcal{D}\psi\bar{\mathcal{D}}\bar{\psi}= \left(\mathcal{D}^{\alpha}\psi_{\beta} \right)\bar{\mathcal{D}}_{\alpha}\bar{\psi}^{\beta}, \quad  \mathcal{D}\bar{\psi}\bar{\mathcal{D}}\psi= \left(\mathcal{D}^{\alpha}\bar{\psi}^{\beta} \right)\bar{\mathcal{D}}_{\alpha}\psi_{\beta}.
\eea
Our convention for other contractions similar to (\ref{abbr}) is that spinor indices entering the first factor on the right hand side stand in their natural position, i.e. $\psi_\alpha$, $\bar\psi^\alpha$, $\mathcal{D}^\alpha$, $\bar{\mathcal{D}}_\alpha$.

The algebra of the covariant derivatives (\ref{cda}) along with the generalised homogeneity conditions (\ref{const}) yield further quadratic restrictions
\bea\label{const3}
&&
\left(\mathcal{D}^{\alpha} \psi_{\gamma}\right)  \mathcal{D}^{\beta} \bar\psi^\gamma+\left(\mathcal{D}^{\beta} \psi_{\gamma}\right)  \mathcal{D}^{\alpha} \bar\psi^\gamma=0, \quad \left(\bar{\mathcal{D}}_{\alpha} \psi_{\gamma}\right) \bar{\mathcal{D}}_{\beta} \bar\psi^\gamma+\left(\bar{\mathcal{D}}_{\beta} \psi_{\gamma}\right) \bar{\mathcal{D}}_{\alpha} \bar\psi^\gamma=0,
\nonumber\\[2pt]
&&
\left(\mathcal{D}^{\alpha} \psi_{\gamma}\right) \bar{\mathcal{D}}_{\beta} \bar\psi^\gamma+\left(\bar{\mathcal{D}}_{\beta} \psi_{\gamma}\right)  \mathcal{D}^{\alpha} \bar\psi^\gamma=
\frac 12 {\delta_\beta}^\alpha \left(\mathcal{D}\psi\bar{\mathcal{D}}\bar\psi+\mathcal{D}\bar{\psi}\bar{\mathcal{D}}\psi \right),
\eea
where Eq. (\ref{const2}) and the identity $\partial_t=\frac 12 \left(\mathcal{D}\psi\bar{\mathcal{D}}\bar\psi + \mathcal{D}\bar{\psi}\bar{\mathcal{D}}\psi \right)\partial'_t+\left(\partial_t \psi_\alpha\right) \mathcal{D}'^{\alpha}+\left(\partial_t \bar\psi^\alpha\right) \bar{\mathcal{D}'}_\alpha $ were used. The explicit solution to these equations will be discussed in Sect. 5.

In what follows, two corollaries of the last equation in (\ref{const3})
\bea\label{const4}
&&
\mathcal{D}^{\alpha}\left(\mathcal{D}\psi\bar{\mathcal{D}}\bar{\psi}+\mathcal{D}\bar{\psi}\bar{\mathcal{D}}\psi\right) = 4i\left(\partial_{t}\psi_{\beta}\mathcal{D}^{\alpha}\bar{\psi}^{\beta} + \partial_{t}\bar{\psi}^{\beta}\mathcal{D}^{\alpha}\psi_{\beta}\right),
\nonumber
\\[2pt]
&&
\bar{\mathcal{D}}_{\alpha}\left(\mathcal{D}\psi\bar{\mathcal{D}}\bar{\psi}+\mathcal{D}\bar{\psi}\bar{\mathcal{D}}\psi\right) = 4i\left(\partial_{t}\psi_{\beta}\bar{\mathcal{D}}_{\alpha}\bar{\psi}^{\beta} + \partial_{t}\bar{\psi}^{\beta}\bar{\mathcal{D}}_{\alpha}\psi_{\beta}\right),
\eea
will be extensively used. The complex conjugation rules
\begin{align}
&
{\left( \mathcal{D}^{\alpha} \psi_\beta \right)}^{*}={\bar{\mathcal{D}}}_\alpha \bar\psi^\beta, && {\left( {\bar{\mathcal{D}}}_\alpha \psi_\beta \right)}^{*}=\mathcal{D}^{\alpha} \bar\psi^\beta,
\nonumber
\end{align}
\begin{align}
&
{\left(  \mathcal{D}^{\alpha} \psi^\beta \right)}^{*}=-{\bar{\mathcal{D}}}_\alpha \bar\psi_\beta, && {\left( \mathcal{D}^{\alpha} \bar\psi_\beta \right)}^{*}=-{\bar{\mathcal{D}}}_\alpha \psi^\beta,
\nonumber\\[2pt]
&
{\left(\mathcal{D}^{\alpha} {\bar{\mathcal{D}}}_\beta \psi_\gamma  \right)}^{*}=-{\bar{\mathcal{D}}}_\alpha \mathcal{D}^\beta \bar\psi^\gamma, && {\left(\mathcal{D}^{\alpha} {\bar{\mathcal{D}}}_\beta \bar\psi^\gamma  \right)}^{*}=-{\bar{\mathcal{D}}}_\alpha \mathcal{D}^\beta \psi_\gamma,
\nonumber\\[2pt]
&
{\left( \mathcal{D}^{\alpha}\rho \right)}^{*}=-{\bar{\mathcal{D}}}_\alpha \rho, && {\left(\mathcal{D}^{\alpha} {\bar{\mathcal{D}}}_\beta \rho  \right)}^{*}=-{\bar{\mathcal{D}}}_\alpha \mathcal{D}^\beta \rho,
\end{align}
where $\psi_\alpha$ is a complex fermionic superfield and $\rho$ is a real bosonic superfield,
will prove helpful as well.

Note that for $a=-1$ the supergroup $D(2,1;a)$ simplifies to $SU(1,1|2)\times SU(2)$, which is easily seen by inspecting the structure relations of the corresponding Lie superalgebra (see Appendix B). In this particular case, the chirality condition $\bar{\mathcal{D}}_{\alpha} \psi_\beta=0$ proves to be compatible with $SU(1,1|2)$ symmetry and the homogeneous transformation law for the covariant derivative $\mathcal{D}^{\alpha} = (\mathcal{D}^{\alpha}\psi_{\beta}){\mathcal{D}'}^{\beta}$. The constraints (\ref{const3}) then reduce to
$\left(\mathcal{D}^{\alpha} \psi_{\gamma}\right) \left( \bar{\mathcal{D}}_{\beta} \bar\psi^\gamma\right)=
\frac 12 {\delta_\beta}^\alpha \left(\mathcal{D}\psi\bar{\mathcal{D}}\bar\psi\right)$, which, in their turn, are equivalent to the linear equation
${\mathcal{D}}^\alpha \psi_\beta+ \bar{\mathcal{D}}^\alpha \bar\psi_\beta=0$ \cite{GK}. For generic values of $a$, $D(2,1;a)$ involves transformations which interchange $\psi_\alpha$ and $\bar\psi_\alpha$. For this reason, the chirality condition would be too restrictive and one is led to deal with the whole set of nonlinear constraints (\ref{const3}).

To summarise, the generalised superconformal diffeomorphism of ${\mathcal{R}}^{1|4}$ is specified by a complex fermionic superfield $\psi_\alpha(t,\theta,\bar\theta)$ obeying the quadratic constraints (\ref{const3}). The bosonic partner $\rho(t,\theta,\bar\theta)$ is determined by (\ref{const2}). Below, $\psi_\alpha(t,\theta,\bar\theta)$ will be identified with the argument of a generalised $D(2,1;a)$ super--Schwarzian derivative.

\vspace{0.5cm}

\noindent
{\bf 3. Infinitesimal $D(2,1;a)$ transformations}\\

\noindent
In this section, we construct infinitesimal $D(2,1;a)$ transformations acting upon the form of the superfields $\rho$ and $\psi_\alpha$ introduced in the preceding section. They will prove useful when verifying the invariance of constraints, to be imposed upon $\psi_\alpha$ later, as well as for analysing the infinitesimal limit of finite $D(2,1;a)$ transformations, which result from setting a generalised $D(2,1;a)$ super--Schwarzian derivative to zero. Technically, it suffices to apply the method of nonlinear realisations \cite{CWZ} to $D(2,1;a)$ and treat $\rho$ and $\psi_\alpha$ as Goldstone superfields associated with the generators of translation and supersymmetry transformation, respectively.

As the first step, one considers the structure relations of Lie superalgebra associated with $D(2,1;a)$ (see Appendix B) as well as the $d=1$, $\mathcal{N}=4$ supersymmetry algebra $\{q_{\alpha},\bar{q}^{\beta}\} = 2h{\delta_{\alpha}}^{\beta}$. Then each generator of the former is accompanied by a Goldstone superfield of the same Grassmann parity, while coordinates of $\mathcal{R}^{1|4}$ superspace are linked to $h$ and $q_\alpha$. Afterwards, the group--theoretic element is introduced
\bea\label{gte}
&&
\tilde{g} = e^{ith}e^{\theta^{\alpha}q_{\alpha} + \bar{\theta}_{\alpha}\bar{q}^{\alpha}} e^{i\rho P}e^{\psi^{\alpha}Q_{\alpha} + \bar{\psi}_{\alpha}\bar{Q}^{\alpha}} e^{\phi^{\alpha} S_{\alpha}+\bar{\phi}_{\alpha}\bar{S}^{\alpha}} e^{i\mu K} e^{i\nu D} e^{i\lambda_{l} \mathcal{J}_{l}} e^{i\left(k_{+} I_{+} + k_{-}I_{-}\right)} e^{i k_{3}I_{3}},
\eea
where $(P,D,K)$ are bosonic generators of translations, dilatations, and special conformal transformations, respectively. $\mathcal{J}_l$, with $l=1,2,3$, generate the $R$--symmetry subalgebra $su(2)$. One more $su(2)$ is realised by $I_{\pm}$, $I_3$, for which the Cartan basis is chosen. $Q_\alpha$ and $S_\alpha$ are fermionic generators of supersymmetry transformations and superconformal boosts, $\bar Q^\a$, $\bar S^\a$ being their Hermitian conjugates. Accordingly, $(\rho,\mu,\nu,\lambda_l,k_{3})$ are real bosonic superfields, $k_{-}$ and $k_{+}$ are complex conjugates of each other, while $(\psi_{\alpha},\bar{\psi}^{\alpha})$ and $(\phi_{\alpha},\bar{\phi}^{\alpha})$ form complex conjugate fermionic pairs. In what follows, $\rho$ and $\psi_\alpha$ are identified with those in the preceding section.

As the next step, one considers the left multiplication by a $D(2,1;a)$ group element $g$
\be\label{TR}
\tilde g'= g\cdot \tilde g, \qquad
g=e^{id P} e^{\epsilon^\alpha Q_\alpha+{\bar\epsilon}_\alpha {\bar Q}^\alpha} e^{\kappa^\alpha S_\alpha+{\bar\kappa}_\alpha {\bar S}^\alpha} e^{i c  K} e^{i b D} e^{i v_l \mathcal{J}_l} e^{i\left(p_{+} I_{+} + p_{-}I_{-}\right)} e^{i p_{3}I_{3}},
\ee
where $(d,c,b,v_l,p_{\pm},p_{3})$ and $(\epsilon_\alpha,\kappa_\alpha)$ are bosonic and fermionic parameters, respectively, with ${\left(p_{-}\right)}^{*}=p_{+}$, ${\left(\epsilon_\alpha\right)}^{*}=\bar\epsilon^\alpha$, ${\left(\kappa_\alpha\right)}^{*}=\bar\kappa^\alpha$, and then repeatedly uses the
the Baker--Campbell--Hausdorff formula
\be\label{ser}
e^{iA}~ T~ e^{-iA}=T+\sum_{n=1}^\infty\frac{i^n}{n!}
\underbrace{[A,[A, \dots [A,T] \dots]]}_{n~\rm times}.
\ee
Focusing on the transformation laws for $\rho$ and $\psi_\alpha$ and regarding the parameters to be infinitesimal, one finally gets (below different transformations are separated by semicolons and vanishing variations are omitted)
\bea\label{tr}
&&
\delta_d\rho=d; \qquad
\delta_b\rho=b\rho, \qquad \delta_b\psi_\alpha=\frac 12 b \psi_\alpha, \qquad \delta_b\bar\psi^\alpha=\frac 12 b \bar\psi^\alpha;
\nonumber\\[2pt]
&&
\delta_c\rho=c \rho^2+\frac{(1+2a)}{2} c \psi^2 \bar\psi^2, \qquad \delta_c\psi_\alpha=c \rho \psi_\alpha-\frac{i(1+2a)}{2} c \psi^2 \bar\psi_\alpha,
\nonumber\\[2pt]
&&
\delta_c\bar\psi^\alpha=c\rho \bar\psi^\alpha-\frac{i(1+2a)}{2} c \bar\psi^2 \psi^\alpha; \qquad
\delta_v \psi_\alpha=\frac{i}{2} v_a {{(\s_a)}_\alpha}^\beta \psi_\beta,
\nonumber\\[2pt]
&&
\delta_v \bar\psi^\alpha=-\frac{i}{2} v_a \bar\psi^\beta {{(\s_a)}_\beta}^\alpha; \qquad
\delta_{p_{+}}\bar\psi^\alpha=-p_{+} \psi^\alpha; \qquad
\delta_{p_{-} } \psi_\alpha=p_{-} \bar\psi_\alpha;
\nonumber\\[2pt]
&&
\delta_{p_3} \psi_\alpha=-\frac{i}{2} p_3 \psi_\alpha, \qquad \delta_{p_3} \bar\psi^\alpha=\frac{i}{2} p_3 \bar\psi^\alpha;
\nonumber\\[2pt]
&&
\delta_\epsilon \rho=i\left(\bar\psi\epsilon-\bar\epsilon\psi\right), \qquad \delta_\epsilon \psi_\alpha=\epsilon_\alpha, \qquad \delta_\epsilon \bar\psi^\alpha=\bar\epsilon^\alpha;
\nonumber\\[10pt]
&&
\delta_\kappa \rho=-\bar\psi\kappa\left(i \rho+(1+2a) \bar\psi\psi \right)+\bar\kappa \psi \left(i \rho-(1+2a) \bar\psi\psi \right),
\nonumber\\[10pt]
&&
\delta_\kappa \psi_\alpha=-\rho \kappa_\alpha+i\bar\psi\psi \kappa_\alpha
-2i(1+a)\kappa_\beta \psi^\beta \bar\psi_\alpha+ia\psi^2 \bar\kappa_\alpha,
\nonumber\\[10pt]
&&
\delta_\kappa \bar\psi^\alpha=-\rho \bar\kappa^\alpha-i\bar\psi\psi \bar\kappa^\alpha-2i(1+a) \bar\kappa^\beta \bar\psi_\beta \psi^\alpha+i a \bar\psi^2 \kappa^\alpha,
\eea
where ${{(\s_a)}_\beta}^\alpha$ are the Pauli matrices (see Appendix A). Above both the original and transformed superfields depend on the same arguments, i.e. $\delta \rho=\rho'(t,\theta,\bar\theta)-\rho(t,\theta,\bar\theta)$, $\delta \psi_\alpha=\psi'_\alpha(t,\theta,\bar\theta)-\psi_\alpha(t,\theta,\bar\theta)$. Representing a given variation $\delta$ as the product of a parameter and a generator, e.g. $\delta_c=-ic K$, $\delta_\epsilon=\epsilon^\alpha Q_\alpha+\bar\epsilon_\alpha \bar Q^\alpha$, and computing commutators $[\delta_1,\delta_2]$ of variations acting upon the superfields $\rho$ and $\psi_\alpha$, one can verify that (\ref{tr}) does reproduce the structure relations of Lie superalgebra associated with $D(2,1;a)$, which are gathered in Appendix B.

\vspace{0.5cm}

\noindent
{\bf 4. A group--theoretic analysis}\\

\noindent
In this section, we extend a recent analysis of $SU(1,1|2)$ \cite{GK} to the case of $D(2,1;a)$. Our primary concern is to construct the analogues of the Maurer--Cartan invariants, which gave rise to the $SU(1,1|2)$ super--Schwarzian derivative in \cite{GK}.

In the previous section, the group--theoretic element (\ref{gte}) was introduced. By making use of the covariant derivative $\mathcal{D}^{\alpha}$, one can build the Grassmann--odd analogues of the Maurer-Cartan one--forms
\bea\label{MC}
&&
\tilde{g}^{-1}\mathcal{D}^{\alpha}\tilde{g} =i {\left(\omega_{P}\right)}^{\alpha} P+ i{\left(\omega_D\right)}^{\alpha} D +i{\left(\omega_K\right)}^{\alpha} K   + i{\left(\omega_{\mathcal{J}}\right)}_b{}^{\alpha}  \mathcal{J}_{b} + i{\left(\omega_{-}\right)}^{\alpha} I_{-} + i{\left(\omega_{+}\right)}^{\alpha} I_{+}
\nonumber
\\[2pt]
&&
\qquad \qquad ~+i{\left(\omega_{3}\right)}^\alpha I_{3}
+ {\left(\omega_{Q}\right)}^{\alpha\beta} Q_{\beta} + {\left(\omega_{\bar{Q}}\right)}_\beta{}^{\alpha} \bar{Q}^{\beta} + {\left(\omega_{S}\right)}^{\alpha\beta} S_{\beta} + {\left(\omega_{\bar{S}} \right)}_\beta{}^{\alpha} \bar{S}^{\beta}-q^\alpha.
\eea
The superfields ${\left(\omega_P\right)}^{\alpha},\dots,{\left(\omega_{\bar{S}} \right)}_\beta{}^{\alpha}$ turn out to be rather bulky and are displayed in Appendix C.
By construction, $\tilde{g}^{-1}\mathcal{D}^{\alpha}\tilde{g}$ hold invariant under the transformation (\ref{TR}) and, hence, ${\left(\omega_P\right)}^{\alpha},\dots,{\left(\omega_{\bar{S}} \right)}_\beta{}^{\alpha}$ provide $D(2,1;a)$ invariants. They can be used to impose constraints which allow one to eliminate some of the superfields entering (\ref{gte}) from the consideration as well as to study candidates for a $D(2,1;a)$ super--Schwarzian derivative. Note that ${\tilde g}^{-1} \bar{\mathcal{D}}_\alpha \tilde g$ results in the complex conjugate invariants.

In a series of recent works \cite{AG,GK,AG2}, it was suggested to choose the constraints so as to link {\it all} the superfields entering the group--theoretic element (\ref{gte}) to $\rho$ and $\psi_\alpha$. Unfortunately, for the case at hand it proved difficult to handle the triplet $(k_{\pm},k_3)$. Below, we follow a broader road and impose just as many constraints as is needed to build an analogue of the $SU(1,1|2)$ super--Schwarzian derivative in \cite{MU,GK}.

All the reservations made, let us discuss the Maurer--Cartan invariants. Comparing ${\left(\omega_{P}\right)}^{\alpha}$ in Appendix C with Eq. (\ref{const1}) above, one concludes that it vanishes.
Had we not chosen to impose (\ref{const1}) earlier, it might have been obtained here by setting ${\left(\omega_{P}\right)}^{\alpha}=0$.

A study of the $D(2,1;-1)$ case in \cite{GK} showed that the fermionic superfield $\psi_\alpha$ was to be chiral. Within the method of nonlinear realisations, a suitable constraint was provided by ${\left(\omega_{\bar{Q}} \right)}_\beta{}^{\alpha}=0$. Imposing a similar condition for generic $a$, one gets
\be\label{const5}
\bar{\mathcal{D}}_{\alpha}\psi_{\beta} = \left(\bar{\mathcal{D}}_{\alpha}\bar{\psi}_{\beta}\right)\frac{k\tan{k}}{k_{+}}, \qquad
\mathcal{D}^{\alpha}\bar{\psi}^{\beta} = -\left(\mathcal{D}^{\alpha}\psi^{\beta}\right)\frac{k\tan{k}}{k_{-}},
\ee
where $k= \sqrt{k_{-}k_{+}}$. These equations allow one to express $\frac{k\tan{k}}{k_{\pm}}$ in terms of $\psi_\alpha$, but more importantly, they imply
an extra (complex) quadratic constraint on the fermionic superfield
\be\label{const6}
\left(\mathcal{D}^{\alpha}\psi_{\beta}\right) \mathcal{D}^{\mu} \bar\psi_{\nu}=
\left(\mathcal{D}^{\mu} \psi_{\nu}\right) \mathcal{D}^{\alpha}\bar{\psi}_{\beta},
\ee
which, in its turn, has an impact on the restrictions (\ref{const3}) revealed above. The first two conditions in (\ref{const3}) are now satisfied identically, while the last equation simplifies to\footnote{Note that (\ref{const5}) and (\ref{const7}) imply $\mathcal{D}^{\alpha} \bar\psi^\gamma \bar{\mathcal{D}}_{\beta} \psi_\gamma=\frac 12 {\delta_\beta}^\alpha \mathcal{D}\bar\psi \bar{\mathcal{D}}\psi$,
which can also be obtained by applying $\delta_{p_{\pm}}$--transformations in (\ref{tr}) to Eq. (\ref{const7}).}
\be\label{const7}
\left(\mathcal{D}^{\alpha} \psi_{\gamma}\right) \bar{\mathcal{D}}_{\beta} \bar\psi^\gamma=
\frac 12 {\delta_\beta}^\alpha \mathcal{D}\psi\bar{\mathcal{D}}\bar\psi,
\ee
which means that, up to a scalar superfield factor, $\mathcal{D}^{\alpha} \psi_{\gamma}$ is a unitary matrix. Given the infinitesimal $D(2,1;a)$ transformations (\ref{tr}), it is straightforward to verify that (\ref{const6}) and (\ref{const7}) do hold invariant.

Recall that for $a=-1$ the superfield $\psi_\alpha$ can be chosen chiral. Hence, (\ref{const6}) is irrelevant in that case, while (\ref{const7}) guarantees that the covariant derivatives algebra $\{\mathcal{D}^{\alpha},\bar{\mathcal{D}}_{\beta} \}=2 i {\delta_\beta}^\alpha \partial_t$ is preserved under the superconformal diffeomorphism $\mathcal{D}^{\alpha} = (\mathcal{D}^{\alpha}\psi_{\beta}){\mathcal{D}'}^{\beta}$. As was mentioned above, the chirality condition turns out to be
incompatible with $D(2,1;a)$ symmetry and the meaning of the novel restriction (\ref{const6}) is that $\mathcal{D}^{\alpha} \bar\psi_{\beta}$ is proportional to $\mathcal{D}^{\alpha} \psi_{\beta}$.

A group--theoretic derivation of the $SU(1,1|2)$ super--Schwarzian derivative in \cite{GK} revealed that it is constructed in terms of the dilaton superfield $\nu$, which, in its turn, is linked to $\psi_\alpha$  by means of constraints. Let us carry out a similar analysis for an arbitrary value of $a$.

As the first step, one expresses $\nu$ in terms of $\psi_\alpha$ by imposing the constraint
\be\label{const8}
{\left(\omega_{Q}\right)}^{\alpha\beta}= r^{\alpha\beta},
\ee
where $r^{\alpha\beta}$ is a constant matrix (composed of coupling constants). Contracting
${\left(\omega_{Q}\right)}^{\alpha\beta}$ with its Hermitian conjugate, one finds
\be\label{Nu}
e^{\nu} = \frac{\mathcal{D}\psi\bar{\mathcal{D}}\bar\psi + \mathcal{D}\bar{\psi}\bar{\mathcal{D}}\psi}{r \bar r},
\ee
where $r \bar r =r^{\alpha\beta}\bar{r}_{\beta\alpha}$.

Then one demands
\be
{\left(\omega_{D}\right)}^{\alpha}=0,
\ee
and takes into account (\ref{const4}) and (\ref{Nu}), which all together link $\phi_\alpha$ to $\psi_\alpha$
\bea\label{phi1}
&&
\phi_{\alpha} = -\frac{2\partial_{t}\psi_{\alpha}}{\mathcal{D}\psi\bar{\mathcal{D}}\bar{\psi}+\mathcal{D}\bar{\psi}\bar{\mathcal{D}}\psi}, \quad \bar{\phi}^{\alpha} = -\frac{2\partial_{t}\bar{\psi}^{\alpha}}{\mathcal{D}\psi\bar{\mathcal{D}}\bar{\psi}+\mathcal{D}\bar{\psi}\bar{\mathcal{D}}\psi}.
\eea

Finally, guided by the previous studies of the $a=-1$ case in \cite{GK}, one considers the symmetric tensor invariant
\be
{{\left(\omega_{S}\right)}_\alpha}^\gamma {\left(\bar{\omega}_{Q}\right)}_{\gamma\beta}+{{\left(\omega_{S}\right)}_\beta}^\gamma {\left(\bar{\omega}_{Q}\right)}_{\gamma\alpha},
\ee
where ${\left(\bar{\omega}_{Q}\right)}_{\beta\alpha}={\left({\left(\omega_{Q}\right)}^{\alpha\beta}\right)}^{*}$. The explicit calculation relates it to
\bea\label{candid}
&&
\mathcal{I}_{\alpha\beta}[\psi(t,\theta,\bar\theta);t,\theta,\bar\theta]:=\mathcal{D}_{\alpha} \bar{\mathcal{D}}_{\beta} \ln{\left( \mathcal{D}\psi\bar{\mathcal{D}}\bar{\psi}+\mathcal{D}\bar{\psi}\bar{\mathcal{D}}\psi\right) }
-2 i \bar\phi^\gamma \mathcal{D}_{\alpha} \bar{\mathcal{D}}_{\beta} \psi_\gamma+2i \phi^\gamma \mathcal{D}_{\alpha} \bar{\mathcal{D}}_{\beta} \bar\psi_\gamma
\nonumber\\[2pt]
&&
+(1+a) \left[ \mathcal{D}_{\alpha} \ln{\left( \mathcal{D}\psi\bar{\mathcal{D}}\bar{\psi}+\mathcal{D}\bar{\psi}\bar{\mathcal{D}}\psi\right) } \right]  \bar{\mathcal{D}}_{\beta} \ln{\left( \mathcal{D}\psi\bar{\mathcal{D}}\bar{\psi}+\mathcal{D}\bar{\psi}\bar{\mathcal{D}}\psi\right) }
+\left(\alpha \leftrightarrow \beta\right),
\eea
which involves $\psi_\alpha$ alone. Note that for $a=-1$ and $\bar{\mathcal{D}}_{\alpha} \psi_\beta=0$ it correctly reproduces the $\mathcal{N}=4$ super--Schwarzian derivative \cite{MU,GK}
\be\label{N4Sch}
\left(\mathcal{D}_{\alpha} \bar{\mathcal{D}}_{\beta} +\mathcal{D}_{\beta} \bar{\mathcal{D}}_{\alpha} \right) \ln{\mathcal{D}\psi\bar{\mathcal{D}}\bar{\psi}}.
\ee

As was mentioned in the Introduction, a super--Schwarzian derivative
$S[\psi(t,\theta);t,\theta]$ must obey two conditions.\footnote{Our presentation here is schematic. External indices carried by a super--Schwarzian derivative and Grassmann--odd coordinates are omitted.} Firstly, considering a superconformal diffeomorphism $t'=\rho(t,\theta)$, $\theta'=\psi(t,\theta)$
and changing the argument $\psi(t,\theta) \to \Omega(t',\theta')$, one should reveal the identity
\be\label{CL}
S[\Omega(t',\theta');t,\theta]=S[\psi(t,\theta);t,\theta]+\mathcal{M}(\theta') S[\Omega(t',\theta');t',\theta'],
\ee
with some superfield coefficient $\mathcal{M}(\theta')$.
Eq. (\ref{CL}) is known as the composition law. Secondly, setting $S[\psi(t,\theta);t,\theta]$ to vanish, one should reproduce a finite--dimensional superconformal transformation acting in the Grassmann--odd sector of superspace. In view of (\ref{CL}), this would determine the symmetry group of $S[\psi(t,\theta);t,\theta]$.

Focusing on the candidate (\ref{candid}) and taking into account the identity
\be\label{ID}
\mathcal{D}\Omega \bar{\mathcal{D}}\bar{\Omega}+\mathcal{D}\bar{\Omega}\bar{\mathcal{D}}\Omega=\frac 12 \left(\mathcal{D}\psi\bar{\mathcal{D}}\bar{\psi}+\mathcal{D}\bar{\psi}\bar{\mathcal{D}}\psi\right)
\left(\mathcal{D}'\Omega \bar{\mathcal{D}}'\bar{\Omega}+\mathcal{D}'\bar{\Omega}\bar{\mathcal{D}}'\Omega\right),
\ee
where (\ref{const}), (\ref{const6}), (\ref{const7}) were used, one finds that it does not obey the composition law (\ref{CL}) unless extra constraints including
\be\label{const9}
\left( \mathcal{D}_{\alpha} \psi^\beta \right) \bar{\mathcal{D}}_{\mu} \bar\psi^\nu=\left( \mathcal{D}_{\mu} \psi^\nu \right) \bar{\mathcal{D}}_{\alpha} \bar\psi^\beta,
\ee
and
\bea\label{const10}
&&
\mathcal{D}_{\alpha} \bar{\mathcal{D}}_{\beta} \psi_\gamma+(1+a) \left[\left(\mathcal{D}_{\alpha} \psi_\gamma \right) \bar{\mathcal{D}}_{\beta} \ln\left(\mathcal{D}\psi\bar{\mathcal{D}}\bar{\psi}+\mathcal{D}\bar{\psi}\bar{\mathcal{D}}\psi\right)
\right.
\nonumber\\[2pt]
&&
\left.
-\left(\bar{\mathcal{D}}_{\beta} \psi_\gamma \right)\mathcal{D}_{\alpha} \ln\left(\mathcal{D}\psi\bar{\mathcal{D}}\bar{\psi}+\mathcal{D}\bar{\psi}\bar{\mathcal{D}}\psi\right)   \right]+\left(\alpha \leftrightarrow \beta\right)=0,
\eea
and
\bea\label{const11}
&&
\mathcal{D}_{\alpha} \bar{\mathcal{D}}_{\beta} \bar\psi_\gamma
+(1+a) \left[\left(\mathcal{D}_{\alpha} \bar\psi_\gamma \right) \bar{\mathcal{D}}_{\beta} \ln\left(\mathcal{D}\psi\bar{\mathcal{D}}\bar{\psi}+\mathcal{D}\bar{\psi}\bar{\mathcal{D}}\psi\right)
\right.
\nonumber\\[2pt]
&&
\left.
-\left(\bar{\mathcal{D}}_{\beta} \bar\psi_\gamma \right)\mathcal{D}_{\alpha} \ln\left(\mathcal{D}\psi\bar{\mathcal{D}}\bar{\psi}+\mathcal{D}\bar{\psi}\bar{\mathcal{D}}\psi\right)   \right]+\left(\alpha \leftrightarrow \beta\right)=0,
\eea
are imposed.\footnote{Note that (\ref{const10}) and (\ref{const11}) could be obtained from a
more general $D(2,1;a)$--invariant constraint
\bea
&&
\mathcal{D}_{\alpha}\bar{\mathcal{D}}_{\beta}\psi_{\gamma} + a
\epsilon_{\alpha\beta} (\mathcal{\bar{D}}_{\mu}\psi_{\gamma})
\mathcal{D}^{\mu}\ln{(\mathcal{D}\psi\bar{\mathcal{D}}\bar{\psi} +
\mathcal{D}\bar{\psi}\bar{\mathcal{D}}\psi)} +
\nonumber
\\[2pt]
&&
+
(1+a)\left[(\mathcal{D}_{\beta}\psi_{\gamma})\bar{\mathcal{D}}_{\alpha}\ln{(\mathcal{D}\psi\bar{\mathcal{D}}\bar{\psi}
+ \mathcal{D}\bar{\psi}\bar{\mathcal{D}}\psi)}  -
(\bar{\mathcal{D}}_{\beta}\psi_{\gamma})\mathcal{D}_{\alpha}\ln{(\mathcal{D}\psi\bar{\mathcal{D}}\bar{\psi}
+ \mathcal{D}\bar{\psi}\bar{\mathcal{D}}\psi)}\right]   = 0
\nonumber
\eea
by symmetrising $\alpha$ and $\beta$.} Taking into account that the infinitesimal transformations (\ref{tr}) leave $\mathcal{D}\psi\bar{\mathcal{D}}\bar{\psi}+\mathcal{D}\bar{\psi}\bar{\mathcal{D}}\psi$ invariant but for the dilatation, special conformal transformation, and superconformal boost
\bea
&&
\mathcal{D}\psi' \bar{\mathcal{D}}\bar{\psi}'+\mathcal{D}\bar{\psi}' \bar{\mathcal{D}}\psi'=(1+b)\left(\mathcal{D}\psi\bar{\mathcal{D}}\bar{\psi}+\mathcal{D}\bar{\psi}\bar{\mathcal{D}}\psi\right),
\nonumber\\[2pt]
&&
\mathcal{D}\psi' \bar{\mathcal{D}}\bar{\psi}'+\mathcal{D}\bar{\psi}' \bar{\mathcal{D}}\psi'=(1+2 c\rho)\left(\mathcal{D}\psi\bar{\mathcal{D}}\bar{\psi}+\mathcal{D}\bar{\psi}\bar{\mathcal{D}}\psi\right),
\nonumber\\[2pt]
&&
\mathcal{D}\psi' \bar{\mathcal{D}}\bar{\psi}'+\mathcal{D}\bar{\psi}' \bar{\mathcal{D}}\psi'=(1+2i(\bar\kappa\psi-\bar\psi\kappa))\left(\mathcal{D}\psi\bar{\mathcal{D}}\bar{\psi}+\mathcal{D}\bar{\psi}\bar{\mathcal{D}}\psi\right),
\eea
one can verify that (\ref{const9}), (\ref{const10}), and (\ref{const11}) do hold invariant. Note that Eq. (\ref{const9}) implies that, up to a scalar superfield factor, $\mathcal{D}_{\alpha} \psi^\beta $ is the same as $\bar{\mathcal{D}}_{\alpha} \bar\psi^\beta$. In particular, the quadratic constraint (\ref{const7}) is now satisfied identically.

In their turn, the restrictions (\ref{const9}), (\ref{const10}), (\ref{const11}) bring (\ref{candid}) to the form
\bea\label{Sch}
&&
\mathcal{I}_{\alpha\beta}[\psi(t,\theta,\bar\theta);t,\theta,\bar\theta]=\mathcal{D}_{\alpha} \bar{\mathcal{D}}_{\beta} \ln{\left( \mathcal{D}\psi\bar{\mathcal{D}}\bar{\psi}+\mathcal{D}\bar{\psi}\bar{\mathcal{D}}\psi\right) }
\nonumber\\[2pt]
&&
\qquad
-(1+a) \left[\mathcal{D}_{\alpha} \ln{\left( \mathcal{D}\psi\bar{\mathcal{D}}\bar{\psi}+\mathcal{D}\bar{\psi}\bar{\mathcal{D}}\psi\right) }\right]  \bar{\mathcal{D}}_{\beta} \ln{\left( \mathcal{D}\psi\bar{\mathcal{D}}\bar{\psi}+\mathcal{D}\bar{\psi}\bar{\mathcal{D}}\psi\right) }
+\left(\alpha \leftrightarrow \beta\right),
\eea
which is characterised by the transition matrix
$\mathcal{M}_{\alpha\beta}^{\mu\nu} (\theta',\bar\theta')=\left(\bar{\mathcal{D}}_{\alpha} \theta'^\mu \right) \mathcal{D}_{\beta} \bar\theta'^\nu-\left(\bar{\mathcal{D}}_{\beta} \bar\theta'^\nu \right) \mathcal{D}_{\alpha} \theta'^\mu$ as far as the composition law (\ref{CL}) is concerned. Let us stress again that both the composition law and $D(2,1;a)$ symmetry of (\ref{Sch}) essentially rely upon the supplementary conditions (\ref{const9}), (\ref{const10}), (\ref{const11}).

To summarise, Eq. (\ref{Sch}), in which $\psi_\alpha$ is assumed to obey (\ref{const6}), (\ref{const9}), (\ref{const10}), appears to be a reasonable generalisation of the $\mathcal{N}=4$ super--Schwarzian derivative (\ref{N4Sch}) to the $D(2,1;a)$ case. In order to establish that (\ref{Sch}) is indeed a legitimate candidate, one has to solve (\ref{const6}), (\ref{const9}), (\ref{const10}) explicitly along with the equation $\mathcal{I}_{\alpha\beta}[\psi(t,\theta,\bar\theta);t,\theta,\bar\theta]=0$ and verify that the resulting superfield $\psi_\alpha$ coincides with a finite form of $D(2,1;a)$ transformation acting in the Grassmann--odd sector of $\mathcal{R}^{1|4}$ superspace. This issue is analysed in the next section.

\vspace{0.5cm}

\noindent
{\bf 5. Solution of constraints}\\

\noindent
Consider a component decomposition of
a generic fermionic superfield defined on $\mathcal{R}^{1|4}$
\bea\label{csf}
&&
\psi_\sigma(t,\theta,\bar\theta)=\alpha_\sigma(t)+\theta_\lambda {b_\sigma}^\lambda (t)+\bar\theta^\lambda c_{\lambda\sigma} (t)+\theta^2 \beta_\sigma (t)+\bar\theta^2 \gamma_\sigma (t)
+\left(\bar\theta \sigma_a \theta \right)\mu_{a \sigma}(t)
\nonumber\\[2pt]
&&
\qquad \qquad \quad
+\bar\theta\theta \xi_\sigma (t)
+\theta^2 \bar\theta^\lambda d_{\lambda\sigma}(t)
+\bar\theta^2 \theta_\lambda {p_\sigma}^\lambda (t)
+\theta^2 \bar\theta^2 \nu_\sigma (t).
\eea
Here Latin letters stand for bosonic components, Greek letters designate fermionic ones, and ${{(\s_a)}_\a}^\b$, $a=1,2,3$, are the Pauli matrices (see Appendix A).
The complex conjugate partner reads
\bea\label{csf1}
&&
\bar\psi^\sigma(t,\theta,\bar\theta)={\bar\alpha}^\sigma(t)+\theta_\lambda {\bar c}^{\sigma\lambda} (t)+{\bar\theta}^\lambda {\bar b_\lambda}{}^\sigma (t)+\theta^2 \bar\gamma^\sigma (t)+\bar\theta^2 \bar\beta^\sigma (t)
+\left(\bar\theta \sigma_a \theta \right) {{\bar\mu}_a}{}^\sigma(t)
\nonumber\\[2pt]
&&
\qquad \qquad \quad
+\bar\theta\theta \bar\xi^\sigma (t)
+\theta^2 \bar\theta^\lambda {{\bar p}_\lambda}{}^\sigma(t)
+\bar\theta^2 \theta_\lambda {\bar d}^{\sigma\lambda} (t)
+\theta^2 \bar\theta^2 \bar\nu^\sigma (t),
\eea
where
\begin{align}
&
{\left( \alpha_\sigma \right)}^{*}=\bar\alpha^\sigma, && {\left( {b_\sigma}^\lambda  \right)}^{*}={\bar b}_\lambda{}^\sigma, && {\left( c_{\lambda\sigma}  \right)}^{*}={\bar c}^{\sigma\lambda}, &&
{\left( \beta_\sigma \right)}^{*}=\bar\beta^\sigma, && {\left( \gamma_\sigma \right)}^{*}=\bar\gamma^\sigma,
\nonumber\\[2pt]
&
{\left( \xi_\sigma \right)}^{*}=\bar\xi^\sigma, && {\left( {p_\sigma}^\lambda  \right)}^{*}={\bar p}_\lambda{}^\sigma, && {\left( d_{\lambda\sigma}  \right)}^{*}={\bar d}^{\sigma\lambda}, &&
{\left( \mu_{a \sigma} \right)}^{*}=\bar\mu_a{}^\sigma, && {\left( \nu_\sigma \right)}^{*}=\bar\nu^\sigma.
\end{align}

In order to solve the superfield constraints (\ref{const6}), (\ref{const9}), (\ref{const10}), we use the covariant projection method, in which components of a superfield are linked to its covariant derivatives evaluated at $\theta_\alpha=\bar\theta^\alpha=0$. For the case at hand, one finds\footnote{A superfield $A(t,\theta,\bar\theta)$ evaluated at $\theta=\bar\theta=0$ is usually designated by the symbol $A|$.}
\begin{align}\label{Comp}
&
\psi_\sigma |=\alpha_\sigma, && \mathcal{D}^{\alpha} \psi_\sigma |={b_\sigma}^\alpha,
\nonumber
\end{align}
\begin{align}
&
{\bar{\mathcal{D}}}_\alpha \psi_\sigma |=c_{\alpha\sigma}, &&
\mathcal{D}^{\alpha} \bar{\mathcal{D}}_{\beta} \psi_\sigma |={{(\s_a)}_\beta}^\alpha \mu_{a \sigma}+ i {\delta_\beta}^\alpha {\dot\alpha}_\sigma+{\delta_\beta}^\alpha \xi_\sigma,
\nonumber\\[2pt]
&
\mathcal{D}^2 \psi_\sigma |=-4 \beta_\sigma, && {\bar{\mathcal{D}}}^2 \psi_\sigma |=-4 \gamma_\sigma,
\nonumber\\[2pt]
&
\mathcal{D}^{\alpha} {\bar{\mathcal{D}}}^2 \psi_\sigma |=-4 {p_\sigma}^\alpha+2 i {{\dot c}^\alpha}{}_\sigma, && \bar{\mathcal{D}}_{\alpha} \mathcal{D}^2 \psi_\sigma |=-4 d_{\alpha\sigma}+2 i {\dot b}_{\sigma\alpha},
\nonumber\\[2pt]
&
\mathcal{D}^2  {\bar{\mathcal{D}}}^2 \psi_\sigma |=16 \nu_\sigma-4 \ddot\alpha_\sigma+8 i \dot\xi_\sigma; && \bar\psi^\sigma |=\bar\alpha^\sigma,
\nonumber\\[2pt]
&
\mathcal{D}^{\alpha} \bar\psi^\sigma |={\bar c}^{\sigma\alpha}, && {\bar{\mathcal{D}}}_\alpha \bar\psi^\sigma |={{\bar b}_\alpha}{}^\sigma,
\nonumber\\[2pt]
&
\mathcal{D}^2 \bar\psi^\sigma |=-4 \bar\gamma^\sigma, &&
\mathcal{D}^{\alpha} \bar{\mathcal{D}}_{\beta} \bar\psi^\sigma |={{(\s_a)}_\beta}^\alpha {{\bar\mu}_a}{}^\sigma+ i {\delta_\beta}^\alpha {\dot{\bar\alpha}}^\sigma+{\delta_\beta}^\alpha \bar\xi^\sigma,
\nonumber\\[2pt]
&
{\bar{\mathcal{D}}}^2 \bar\psi^\sigma |=-4 \bar\beta^\sigma, && \mathcal{D}^{\alpha} {\bar{\mathcal{D}}}^2 \bar\psi^\sigma |=-4 {\bar d}^{\sigma\alpha}+2 i {\dot{\bar b}}{}^{\alpha\sigma},
\nonumber\\[2pt]
&
\bar{\mathcal{D}}_{\alpha} \mathcal{D}^2 \bar\psi^\sigma |=-4 {\bar p}_\alpha{}^\sigma+2i {\dot{\bar c}}{}^\sigma{}_\alpha, &&
\mathcal{D}^2  {\bar{\mathcal{D}}}^2 \bar\psi^\sigma |=16 \bar\nu^\sigma-4 {\ddot{\bar\alpha}}{}^\sigma+8 i {\dot{\bar\xi}}{}^\sigma,
\end{align}
where ${\mathcal{D}}^2={\mathcal{D}}^\alpha {\mathcal{D}}_\alpha$, ${\bar{\mathcal{D}}}^2={\bar{\mathcal{D}}}_\alpha {\bar{\mathcal{D}}}^\alpha$ and
the identities ${\mathcal{D}}^\alpha {\mathcal{D}}^\beta=-\frac 12 \epsilon^{\alpha\beta} {\mathcal{D}}^2$,
${\bar{\mathcal{D}}}_\alpha {\bar{\mathcal{D}}}_\beta=-\frac 12 \epsilon_{\alpha\beta} {\bar{\mathcal{D}}}^2$ were used.

Computing the covariant derivatives of (\ref{const6}), (\ref{const9}), (\ref{const10}) and taking into account (\ref{Comp}), after rather tedious calculation, one gets
\begin{align}\label{CompFin}
&
{b_\sigma}^\lambda (t)=u(t) e^{i v}  {\left( \exp{\left[\frac{i}{2} w_{a}\sigma_{a}\right]}\right)}_\sigma{}^\lambda, && {\bar b}_\lambda{}^\sigma (t)=u(t) e^{-i v}  {\left( \exp{\left[-\frac{i}{2} w_{a}\sigma_{a}\right]}\right)}_\lambda{}^\sigma,
\nonumber\\[2pt]
&
c_{\lambda\sigma} (t)=-{\bar q} {\bar b}_{\lambda\sigma} (t), && {\bar c}^{\sigma\lambda} (t)=q b^{\sigma\lambda} (t),
\nonumber\\[2pt]
&
d_{\lambda\sigma} (t)=\frac{i (1+2a) \dot u (t)}{2 u(t)} b_{\sigma\lambda} (t), && {\bar d}^{\sigma\lambda} (t)=\frac{i (1+2a) \dot u (t)}{2 u(t)} {\bar b}^{\lambda\sigma} (t),
\nonumber\\[2pt]
&
{p_\sigma}^\lambda (t)=-\bar q {\bar d}_\sigma{}^\lambda (t), && {\bar p}_\lambda{}^\sigma (t)=q {d_\lambda}^\sigma (t),
\nonumber\\[2pt]
&
\beta_\sigma (t)=\frac{i a e^{2iv}}{1+q \bar q} \left(q \dot\alpha_\sigma (t)-\dot{\bar\alpha}_\sigma (t) \right), && {\bar\beta}^\sigma (t)=\frac{-i a e^{-2iv}}{1+q \bar q} \left({\dot\alpha}^\sigma (t)+\bar q {\dot{\bar\alpha}}^\sigma (t) \right),
\nonumber\\[2pt]
&
\gamma_\sigma (t)=-\bar q {\bar\beta}_\sigma (t), && {\bar\gamma}^\sigma (t)=q \beta^\sigma (t),
\nonumber\\[2pt]
&
\xi_\sigma (t)=\frac{i a(1-q\bar q)}{1+q \bar q} {\dot\alpha}_\sigma (t)+\frac{2ia\bar q}{1+q \bar q} {\dot{\bar\alpha}}_\sigma (t), && \bar\xi^\sigma (t)=-\frac{ia(1-q\bar q)}{1+q\bar q} {{\dot{\bar\alpha}}}^\sigma (t)+\frac{2iaq}{1+q\bar q} {\dot\alpha}^\sigma (t),
\nonumber\\[2pt]
&
\mu_{a \sigma} (t)=-\frac{i(1+a)}{u^2 (t)}  {{\left( b(t) \sigma_a \bar b (t) \right)}_\sigma}^\rho \dot\alpha_\rho (t), && {\bar\mu}_a{}^\sigma (t)=\frac{i(1+a)}{u^2 (t)} {\dot{\bar\alpha}}^\rho (t)  {{\left( b(t) \sigma_a \bar b (t) \right)}_\rho}^\sigma,
\nonumber\\[2pt]
&
\nu_\sigma (t)=\frac{(1+2a)}{4} \ddot\alpha_\sigma (t), && \bar\nu^\sigma (t)=\frac{(1+2a)}{4} {\ddot{\bar\alpha}}^\sigma (t).
\end{align}
Here $w_a$ is a real vector parameter generating $SU(2)$ group. Constant parameters $(v,q,\bar q)$, of which $v$ is real while $q$ and $\bar q$ are complex conjugate to each other, give rise to another $SU(2)$.

Note that in obtaining Eq. (\ref{CompFin}), the following identities
\bea
&&
{{\left(\mbox{exp} \left[-\frac{i}{2} w_a \sigma_a \right] \right)}_\alpha}^\gamma \epsilon_{\gamma\beta}=-{{\left(\mbox{exp} \left[\frac{i}{2} w_a \sigma_a\right] \right)}_\beta}^\gamma \epsilon_{\gamma\alpha},
\nonumber
\eea
\bea
&&
\epsilon^{\alpha\gamma} {{\left(\mbox{exp} \left[-\frac{i}{2} w_a \sigma_a \right] \right)}_\gamma}^\beta =-\epsilon^{\beta\gamma} {{\left(\mbox{exp} \left[\frac{i}{2} w_a \sigma_a\right] \right)}_\gamma}^\alpha,
\nonumber
\eea
were repeatedly used.

For a real bosonic function $u(t)$ entering (\ref{CompFin}) one reveals the differential equation
\be\label{uu}
u \ddot u-2 {\dot u}^2=0 \qquad \Rightarrow \qquad u(t)=\frac{1}{c_0+c_1 t},
\ee
where $c_0$ and $c_1$ are constants of integration. Along with an additive constant of integration, which occurs when solving (\ref{const2}), $c_0$ and $c_1$ generate $SL(2,R)$ transformations.

The complex fermionic function $\alpha_\sigma (t)$ in (\ref{CompFin}) is found to obey two differential equations
\be\label{aa}
u {\ddot\alpha}_\sigma-2 \dot u {\dot\alpha}_\sigma=0, \qquad \dot\alpha_\sigma {\dot{\bar\alpha}}_\lambda+\dot\alpha_\lambda {\dot{\bar\alpha}}_\sigma=0
\qquad \Rightarrow \qquad \alpha_\sigma(t)=\epsilon_\sigma+\frac{(\bar\kappa \kappa) \kappa_\sigma}{c_0+c_1 t},
\ee
where the Grassmann--odd parameters $\epsilon_\sigma$ and $\kappa_\sigma$ are associated with the global supersymmetry transformations and superconformal boosts, respectively.

The resulting superfield (\ref{csf}) determines a finite $D(2,1;a)$ transformation acting in the Grassmann--odd sector of $\mathcal{R}^{1|4}$ superspace, which correctly reduces to (\ref{tr}) in the infinitesimal limit. In particular, in order to reproduce the infinitesimal form of the superconformal boosts entering (\ref{tr}), one sets $c_0=1$, considers $c_1$ to be small, such that $\frac{1}{1+c_1 t}\approx  1-c_1 t$, and identifies $c_1 (\bar\kappa \kappa) \kappa_\gamma$ with the infinitesimal $\kappa_\gamma$ in (\ref{tr}). The resulting transformation is a superposition of the supersymmetry transformation, special conformal transformation parametrized by $c_1$ and the superconformal boost associated with $c_1 (\bar\kappa \kappa) \kappa_\gamma$. A finite $D(2,1;a)$ transformation acting in the Grassmann--even sector of $\mathcal{R}^{1|4}$ can be found by integrating (\ref{const2}).

That a finite form of $D(2,1;a)$ transformation showed up prior to setting (\ref{Sch}) to zero turns out to be a troubling news. For $\mathcal{N}=2,3,4$ super--Schwarzian derivatives, the solution of supplementary conditions similar to (\ref{const6}) and (\ref{const9}) leaves one with an infinite--dimensional supergroup \cite{AG,GK,AG2}, while setting a super--Schwarzian derivative to vanish reduces it to a finite--dimensional superconformal subgroup.
It appears that the extra nonlinear restriction (\ref{const10}) supersedes the equation $\mathcal{I}_{\alpha\beta}[\psi(t,\theta,\bar\theta);t,\theta,\bar\theta]=0$. Indeed, by analysing the latter condition, one reproduces the equations (\ref{uu}) and (\ref{aa}) and, hence, $\mathcal{I}_{\alpha\beta}[\psi(t,\theta,\bar\theta);t,\theta,\bar\theta]$ vanishes identically, provided (\ref{const6}), (\ref{const9}), (\ref{const10}) are satisfied. We tried to prove that (\ref{Sch}) can be algebraically deduced from (\ref{const6}), (\ref{const9}), (\ref{const10}) but failed.

It has to be kept in mind, however, that for $a=-1$ and chiral $\psi_\alpha$, the extra condition (\ref{const10}) vanishes and (\ref{Sch}) comes onto the scene as the legitimate $SU(1,1|2)$ super--Schwarzian derivative.

\vspace{0.5cm}

\noindent
{\bf 6. An alternative candidate}\\

\noindent
That (\ref{Sch}) is an unsuitable candidate for a $D(2,1;a)$ super--Schwarzian derivative comes as a nasty surprise, which forces us to take a step back and reconsider the material in Sect. 4.

Let $\psi_\alpha$ be a complex fermionic superfield, which is subject to two quadratic constraints
\be\label{constFin}
\left(\mathcal{D}^{\alpha}\psi_{\beta}\right) \mathcal{D}^{\mu} \bar\psi_{\nu}=
\left(\mathcal{D}^{\mu} \psi_{\nu}\right) \mathcal{D}^{\alpha}\bar{\psi}_{\beta}, \qquad \left( \mathcal{D}_{\alpha} \psi^\beta \right) \bar{\mathcal{D}}_{\mu} \bar\psi^\nu=\left( \mathcal{D}_{\mu} \psi^\nu \right) \bar{\mathcal{D}}_{\alpha} \bar\psi^\beta.
\ee
As was mentioned above, the former is the analogue of the chirality condition, while the latter guarantees that the covariant derivatives algebra is preserved under the generalised superconformal diffeomorphism. Our analysis in the preceding section shows that, although Eqs. (\ref{constFin}) relate some of the components in (\ref{csf}) to each other, they do not fix $u(t)$ and $\alpha_\sigma (t)$ and, hence, leaves one with an
infinite--dimensional group of transformations.

Let us introduce the third--rank tensor
\bea\label{SchFin}
&&
S_{(\alpha\beta)\gamma}[\psi(t,\theta,\bar\theta);t,\theta,\bar\theta]=
\mathcal{D}_{\alpha} \bar{\mathcal{D}}_{\beta} \psi_\gamma+(1+a) \left[\left(\mathcal{D}_{\alpha} \psi_\gamma \right) \bar{\mathcal{D}}_{\beta} \ln\left(\mathcal{D}\psi\bar{\mathcal{D}}\bar{\psi}+\mathcal{D}\bar{\psi}\bar{\mathcal{D}}\psi\right)
\right.
\nonumber\\[2pt]
&&
\left.
\qquad \qquad \qquad \qquad \qquad
-\left(\bar{\mathcal{D}}_{\beta} \psi_\gamma \right)\mathcal{D}_{\alpha} \ln\left(\mathcal{D}\psi\bar{\mathcal{D}}\bar{\psi}+\mathcal{D}\bar{\psi}\bar{\mathcal{D}}\psi\right)   \right]+\left(\alpha \leftrightarrow \beta\right),
\eea
which is symmetric in the first pair of indices,
and its complex conjugate partner
\bea\label{SchFin1}
&&
{\bar S}_{(\alpha\beta)\gamma}[\psi(t,\theta,\bar\theta);t,\theta,\bar\theta]=
\mathcal{D}_{\alpha} \bar{\mathcal{D}}_{\beta} \bar\psi_\gamma
+(1+a) \left[\left(\mathcal{D}_{\alpha} \bar\psi_\gamma \right) \bar{\mathcal{D}}_{\beta} \ln\left(\mathcal{D}\psi\bar{\mathcal{D}}\bar{\psi}+\mathcal{D}\bar{\psi}\bar{\mathcal{D}}\psi\right)
\right.
\nonumber\\[2pt]
&&
\left.
\qquad \qquad \qquad \qquad \qquad
-\left(\bar{\mathcal{D}}_{\beta} \bar\psi_\gamma \right)\mathcal{D}_{\alpha} \ln\left(\mathcal{D}\psi\bar{\mathcal{D}}\bar{\psi}+\mathcal{D}\bar{\psi}\bar{\mathcal{D}}\psi\right)   \right]+\left(\alpha \leftrightarrow \beta\right).
\eea
As follows from our analysis in the preceding section, setting $S_{(\alpha\beta)\gamma}$ to vanish one reduces the infinite--dimensional group of superconformal isomorphisms
to the finite--dimensional subgroup $D(2,1;a)$.

Considering the superconformal diffeomorphism $t'=\rho(t,\theta,\bar\theta)$, $\theta'_\alpha=\psi_\alpha (t,\theta,\bar\theta)$, where $\rho$ and $\psi_\alpha$ are assumed to obey (\ref{const1}) and (\ref{constFin}), respectively, changing the argument $\psi_\alpha (t,\theta,\bar\theta)\to \Omega_\alpha (t',\theta',\bar\theta')$, and
taking into account the identity (\ref{ID}), one obtains the generalised composition law 
\bea\label{SchFin3}
&&
S_{(\alpha\beta)\gamma}[\Omega(t',\theta',\bar\theta');t,\theta,\bar\theta]=
\\[2pt]
&&
\qquad \qquad
=
\left(\mathcal{D}'^\nu \Omega_\gamma \right) S_{(\alpha\beta)\nu}[\psi(t,\theta,\bar\theta);t,\theta,\bar\theta]
+
\left({\bar{\mathcal{D}}'}_\nu \Omega_\gamma \right) {\bar S}_{(\alpha\beta)}{}^\nu[\psi(t,\theta,\bar\theta);t,\theta,\bar\theta]
\nonumber\\[2pt]
&&
\qquad \qquad
-\frac 12 \left[\left(\mathcal{D}_{\alpha} \psi^\mu \right)\bar{\mathcal{D}}_{\beta} \bar\psi^\nu-\left(\bar{\mathcal{D}}_\alpha \psi^\mu \right)\mathcal{D}_\beta \bar\psi^\nu+ \left(\alpha \leftrightarrow \beta\right) \right]
S_{(\mu\nu)\gamma}[\Omega(t',\theta',\bar\theta');t',\theta',\bar\theta'].
\nonumber
\eea
A similar expression holds for the complex conjugate partner
\bea\label{SchFin4}
&&
{\bar S}_{(\alpha\beta)\gamma}[\Omega(t',\theta',\bar\theta');t,\theta,\bar\theta]=
\\[2pt]
&&
\qquad \qquad
=
\left(\mathcal{D}'^\nu {\bar\Omega}_\gamma \right) S_{(\alpha\beta)\nu}[\psi(t,\theta,\bar\theta);t,\theta,\bar\theta]
+
\left({\bar{\mathcal{D}}'}_\nu {\bar\Omega}_\gamma \right) {\bar S}_{(\alpha\beta)}{}^\nu[\psi(t,\theta,\bar\theta);t,\theta,\bar\theta]
\nonumber\\[2pt]
&&
\qquad \qquad
-\frac 12 \left[\left(\mathcal{D}_{\alpha} \psi^\mu \right)\bar{\mathcal{D}}_{\beta} \bar\psi^\nu-\left(\bar{\mathcal{D}}_\alpha \psi^\mu \right)\mathcal{D}_\beta \bar\psi^\nu+ \left(\alpha \leftrightarrow \beta\right) \right]
{\bar S}_{(\mu\nu)\gamma}[\Omega(t',\theta',\bar\theta');t',\theta',\bar\theta'].
\nonumber
\eea
In comparison to the conventional composition law (\ref{CL}),
an unusual feature of (\ref{SchFin3}), (\ref{SchFin4}) is that both $S_{(\alpha\beta)\gamma}$ and ${\bar S}_{(\alpha\beta)\gamma}$ contribute to the right hand side.

Another unconventional feature of $S_{(\alpha\beta)\gamma}$ is that it does not hold exactly invariant under a finite $D(2,1;a)$ transformation acting upon the argument $\psi_\alpha (t,\theta,\bar\theta)\to \psi'_\alpha(t',\theta',\bar\theta')$. Rather, it transforms covariantly
\bea\label{mis}
&&
S_{(\alpha\beta)\gamma}[\psi'(t',\theta',\bar\theta');t,\theta,\bar\theta]=
\nonumber\\[2pt]
&&
\qquad \qquad
=
\left(\mathcal{D}'^\nu \psi'_\gamma \right) S_{(\alpha\beta)\nu}[\psi(t,\theta,\bar\theta);t,\theta,\bar\theta]
+
\left({\bar{\mathcal{D}}'}_\nu \psi'_\gamma \right) {\bar S}_{(\alpha\beta)}{}^\nu[\psi(t,\theta,\bar\theta);t,\theta,\bar\theta].
\eea
This is an immediate consequence of (\ref{SchFin}) and the fact that $S_{(\alpha\beta)\gamma}$ vanishes when acting upon the fermionic superfield, which determines a $D(2,1;a)$ transformation in the Grassmann--odd sector of $\mathcal{R}^{1|4}$.

Note that (\ref{mis}) resembles the transformation law of the covariant derivative under the generalised superconformal diffeomorphism (\ref{const}). It is rather likely that the difficulty in defining a $D(2,1;a)$ super--Schwarzian derivative with conventional properties is related to the fact that the chirality condition on the fermionic superfield $\psi_\alpha$ is incompatible with $D(2,1;a)$ symmetry. Above, it was this point which led us to admit the mixed transformation rule (\ref{const}).

\vspace{0.5cm}

\noindent
{\bf 7. A parallel with the $\mathcal{N}=3$ case}\\

\noindent
In a recent work \cite{AG2}, the $\mathcal{N}=3$ super--Schwarzian derivative \cite{Sch} was constructed by applying the method of nonlinear realisations to the finite--dimensional superconformal group $OSp(3|2)$.
In this section, we discuss an $\mathcal{N}=3$ analogue of the generalised super--Schwarzian derivative formulated in the preceding section. 

Similarly to the material in Sect. 2, a superconformal diffeomorphism of $\mathcal{R}^{1|3}$ is determined by a real bosonic superfield $\rho$ and a triplet of real fermionic superfields $\psi_i$, $i=1,2,3$, which
give rise to a coordinate transformation $t'=\rho(t,\theta)$, $\theta'_i=\psi_i (t,\theta)$, under which the covariant derivative ${\mathcal{D}}_i=\frac{\vec\partial}{\partial\theta_{i}}-{\rm i} \theta_i \frac{\partial}{\partial t}$ transforms homogeneously ${\mathcal{D}}_i=\left({\mathcal{D}}_i \psi_j\right) {\mathcal{D}}'_j$. The latter condition and the fact that the algebra of covariant derivatives $\{{\mathcal{D}}_i, {\mathcal{D}}_j \}=-2 {\rm i} \delta_{ij} \frac{\partial}{\partial t}$ is preserved, yield constraints upon $\rho$ and $\psi_i$ (for more details see \cite{AG2})
\begin{align}\label{CONST1}
&
{\mathcal{D}}_i \rho+{\rm i} \psi_j {\mathcal{D}}_i \psi_j=0, && \left( {\mathcal{D}}_i \psi_k \right) \left({\mathcal{D}}_j \psi_k\right)=\frac 13 \delta_{ij} {\mathcal{D}}\psi {\mathcal{D}}\psi \qquad \Rightarrow
\nonumber\\[2pt]
&
{\mathcal{D}}_i \left({\mathcal{D}}\psi {\mathcal{D}}\psi \right)=-6 {\rm i} {\dot\psi}_j {\mathcal{D}}_i  \psi_j, && \dot\rho={\rm i} \psi_i {\dot\psi}_i+\frac 13 {\mathcal{D}}\psi {\mathcal{D}}\psi,
\end{align}
where the dot designates the derivative with respect to $t$ and ${\mathcal{D}}\psi {\mathcal{D}}\psi=\left({\mathcal{D}}_i \psi_j \right) \left( {\mathcal{D}}_i \psi_j \right)$.

An $\mathcal{N}=3$ analogue of (\ref{SchFin}) reads
\bea\label{preS}
&&
S_{[ij]k} [\psi(t,\theta);t,\theta]= {\mathcal{D}}_i {\mathcal{D}}_j \psi_k+\left( {\mathcal{D}}_i \psi_k \right) {\mathcal{D}}_j \ln{\left( {\mathcal{D}}\psi {\mathcal{D}}\psi\right)}-\left(i \leftrightarrow j\right),
\eea
which is antisymmetric in the first pair of indices. Considering a superconformal diffeomorphism $t'=\rho(t,\theta)$, $\theta'_i=\psi_i (t,\theta)$ and changing the argument
$\psi_i (t,\theta) \to \Omega_i (t',\theta')$, one finds the generalised composition law
\be\label{CLAW}
S_{[ij]k} [\Omega(t',\theta');t,\theta]=\left( \mathcal{D}'_p \Omega_k \right)  S_{[ij]p} [\psi(t,\theta);t,\theta]+\left( {\mathcal{D}}_i \psi_l \right) \left( {\mathcal{D}}_j \psi_p \right)  S_{[lp]k} [\Omega(t',\theta');t',\theta'],
\ee
where the identity $\left(\mathcal{D} \Omega \mathcal{D} \Omega \right)=\frac 13 \left(\mathcal{D} \psi \mathcal{D} \psi \right) \left(\mathcal{D}' \Omega \mathcal{D}' \Omega \right)$ was used.

Note that Eq. (\ref{preS}) can be obtained by analogy with the $D(2,1;a)$ case. Constructing the Maurer--Cartan invariants similar to those in Appendix C and specifying to ${\left(\omega_S \right)}_{ik} {\left(\omega_Q \right)}_{jk}-{\left(\omega_S \right)}_{jk} {\left(\omega_Q \right)}_{ik}$ one obtains a candidate for an $\mathcal{N}=3$ super--Schwarzian derivative (for more details see \cite{AG2})
\be
[\mathcal{D}_i,\mathcal{D}_j] \ln{(\mathcal{D} \psi \mathcal{D} \psi)}+(\mathcal{D}_i \ln{(\mathcal{D} \psi \mathcal{D} \psi)}) (\mathcal{D}_j \ln{(\mathcal{D} \psi \mathcal{D} \psi)})-\frac{6 {\rm i} {\dot\psi}_l [\mathcal{D}_i,\mathcal{D}_j] \psi_l }{\mathcal{D} \psi \mathcal{D} \psi}.
\ee
Then one reveals that the latter does not obey a composition law unless the constraint ${\mathcal{D}}_i {\mathcal{D}}_j \psi_k+\left( {\mathcal{D}}_i \psi_k \right) {\mathcal{D}}_j \ln{\left( {\mathcal{D}}\psi {\mathcal{D}}\psi\right)}-\left(i \leftrightarrow j\right)=0$ is imposed.

It remains to analyse the equation $S_{[ij]k} [\psi(t,\theta);t,\theta]=0$ and verify that the resulting $\psi_i$ does specify a finite $OSp(3|2)$ transformation acting in the Grassmann--odd sector of $\mathcal{R}^{1|3}$.
Considering a component decomposition
\be\label{ccdec}
\psi_i (t,\theta)=\alpha_i (t)+ \theta_a b_{a i} (t)+\frac 12 \theta_a \theta_b \beta_{a b i} (t)+\frac{1}{3!} \epsilon_{abc} \theta_a \theta_b \theta_c g_i (t),
\ee
where ($b_{a i}$, $g_i$) are Grassmann--even functions of $t$, ($\alpha_i$, $\beta_{a b i}$) are their Grassmann--odd partners, and $\epsilon_{abc}$ is the Levi-Civita symbol, one first computes the covariant projections\footnote{A relation $\epsilon_{qsp} \mathcal{D}_q \mathcal{D}_s \mathcal{D}_p \mathcal{D}_j \mathcal{D}_k \psi_a=6 \epsilon_{jkl} {\ddot b}_{la}$, which holds provided $g_i=0$, proves helpful as well.}
\bea
&&
\psi_i |=\alpha_i, \qquad \mathcal{D}_i \psi_j |=b_{ij}, \qquad \mathcal{D}_i \mathcal{D}_j \psi_k |=-\beta_{ijk}-{\rm i} \delta_{ij} {\dot\alpha}_k,
\nonumber\\[2pt]
&&
\mathcal{D}_i \mathcal{D}_j \mathcal{D}_k \psi_p |=-{\rm i} \delta_{ij} {\dot b}_{kp}-{\rm i} \delta_{jk} {\dot b}_{ip}+{\rm i} \delta_{ik} {\dot b}_{jp}-\epsilon_{ijk} g_p,
\nonumber\\[2pt]
&&
\mathcal{D}_i \mathcal{D}_j \mathcal{D}_k \mathcal{D}_p \psi_l |=-\delta_{ij} \delta_{kp} {\ddot\alpha}_l-\delta_{jk} \delta_{ip} {\ddot\alpha}_l+\delta_{ik} \delta_{jp} {\ddot\alpha}_l-{\rm i} \delta_{ij} {\dot\beta}_{pkl}
-{\rm i} \delta_{jk} {\dot\beta}_{pil}
\nonumber\\[2pt]
&&
\qquad \qquad \qquad \quad ~
+{\rm i} \delta_{ik} {\dot\beta}_{pjl} -{\rm i} \delta_{kp} {\dot\beta}_{jil} +{\rm i} \delta_{jp} {\dot\beta}_{kil} -{\rm i} \delta_{ip} {\dot\beta}_{kjl}.
\eea
Then one solves the quadratic constraint $\left({\mathcal{D}}_j \psi_k\right)=\frac 13 \delta_{ij} {\mathcal{D}}\psi {\mathcal{D}}\psi$ in (\ref{CONST1}), which yields \cite{AG2}
\bea\label{sf1}
b_{ij}=u(t) {\mbox{exp} (\tilde\xi)}_{ij}, \qquad \beta_{ijk}=\frac{3 {\rm i} }{b^2} \left( (b_{is} {\dot\alpha}_s) b_{jk}-(b_{js} {\dot\alpha}_s) b_{ik}\right), \qquad
g_i=\frac{1}{2 u(t)} \epsilon_{ijk} {\dot\alpha}_j {\dot\alpha}_k,
\eea
where $u(t)$ is an arbitrary bosonic function of $t$, the matrix ${\tilde\xi}_{ij}=\xi_k \epsilon_{kij}$ involves a real bosonic vector parameter $\xi_k$ such that ${\mbox{exp} (\tilde\xi)}_{ij}={\mbox{exp} (-\tilde\xi)}_{ji}$, and $b^2=b_{ij} b_{ij}$. Note that, similarly to ${\mathcal{D}}_i \psi_j$, the bosonic component $b_{ij}$ obeys the equation $b_{ik} b_{jk}=\frac 13 \delta_{ij} b^2$, which means that the parameter $\xi_k$ represents a finite $SO(3)$--transformation.

Finally, one sets $S_{[ij]k} [\psi(t,\theta);t,\theta]$ to vanish, which results in the differential equations
\bea
&&
{\dot\alpha}_i {\dot\alpha}_j=0, \qquad u {\ddot\alpha}_i-2 {\dot u } {\dot\alpha}_i=0, \qquad u {\ddot u}-2 {\dot u}^2=0 \qquad \Rightarrow
\nonumber\\[2pt]
&&
u(t)=\frac{1}{c_0+c_1 t}, \qquad \alpha_i (t)=\epsilon_i+\frac{{\rm i} (\epsilon_l \kappa_l) \kappa_i}{c_0+c_1 t},
\eea
where $(c_0,c_1)$ and $(\epsilon_i,\kappa_i)$ are bosonic and fermionic constants of integration, respectively.
The resulting $\psi_i$ precisely coincides with a finite $OSp(3|2)$ transformation acting in the Grassmann--odd sector of $\mathcal{R}^{1|3}$ \cite{AG2}.
Thus, Eq. (\ref{preS}) possesses all the properties of the generalised super--Schwarzian derivative introduced in the preceding section.

Interestingly enough, $S_{[ij]k} [\psi(t,\theta);t,\theta]$ can be regarded as a pre--super--Schwarzian derivative because the conventional $\mathcal{N}=3$ derivative \cite{Sch} can be constructed from it. Indeed, contracting $S_{[ij]k} [\psi(t,\theta);t,\theta]$ with $\frac{{\mathcal{D}}_p \psi_k}{\mathcal{D} \psi \mathcal{D} \psi}$ and performing a cyclic permutation of indices $(p,i,j)$, one gets a totally antisymmetric expression, the only nontrivial component of which reads
\be\label{N3S}
\mathcal{S}[\psi(t,\theta);t,\theta]=\frac{\epsilon_{ijk} (\mathcal{D}_i \psi_l) ( \mathcal{D}_j \mathcal{D}_k \psi_l)}{\mathcal{D} \psi \mathcal{D} \psi}.
\ee
Eq. (\ref{N3S}) is the $\mathcal{N}=3$ super--Schwarzian derivative introduced in \cite{Sch}.

Note that, while (\ref{N3S}) gives a natural Grasmann--odd $OSp(3|2)$ invariant, the generalised object (\ref{preS}) allows one to construct its Grassmann--even counterpart
\be\label{INV}
\frac{S_{[ij]k} [\psi(t,\theta);t,\theta] S_{[ij]k} [\psi(t,\theta);t,\theta]}{\mathcal{D} \psi \mathcal{D} \psi}.
\ee
Indeed, for an $OSp(3|2)$ transformation $\psi_i(t,\theta) \to \psi'_i(t',\theta')$, which acts in the Grassmann--odd sector of $\mathcal{R}^{1|3}$,
the last term in (\ref{CLAW}) vanishes and (\ref{INV}) holds invariant as a consequence of (\ref{CLAW}) and the equalities $\left( {\mathcal{D}}'_i \psi'_k \right) \left({\mathcal{D}}'_j \psi'_k\right)=\frac 13 \delta_{ij} {\mathcal{D}}' \psi' {\mathcal{D}}' \psi'$,
$\left(\mathcal{D} \psi' \mathcal{D} \psi' \right)=\frac 13 \left(\mathcal{D} \psi \mathcal{D} \psi \right) \left(\mathcal{D}' \psi' \mathcal{D}' \psi' \right)$. Because the integration measure in $\mathcal{R}^{1|3}$ is Grassmann--odd, it seems problematic to use (\ref{INV}) within the context of an $\mathcal{N}=3$ Sachdev--Ye--Kitaev theory. Yet, setting (\ref{INV}) to take a (coupling) constant value, one gets a reasonable example of an $\mathcal{N}=3$ super--Schwarzian mechanics, which can be studied along the lines in \cite{AG3}.

Note that, guided by the $\mathcal{N}=3$ analogy, one could try to contract $S_{(\alpha\beta)\gamma}$ and ${\bar S}_{(\alpha\beta)\gamma}$ in the preceding section with the covariant derivatives of $\psi_\gamma$ and $\bar\psi^\gamma$ in an attempt to form a $D(2,1;a)$ super--Schwarzian derivative, obeying the conventional composition law (\ref{CL}). A close inspection shows that, while the expression
\be
\frac{\left(a \bar{\mathcal{D}}_\lambda \bar\psi^\gamma +b \mathcal{D}_\lambda \bar\psi^\gamma  \right) S_{(\alpha\beta)\gamma}+
\left(a \bar{\mathcal{D}}_\lambda \psi_\gamma +b \mathcal{D}_\lambda \psi_\gamma  \right) {\bar S}_{(\alpha\beta)}{}^\gamma}{\mathcal{D}\psi\bar{\mathcal{D}}\bar{\psi}+\mathcal{D}\bar{\psi}\bar{\mathcal{D}}\psi},
\ee
where $a$, $b$ are constants, does hold invariant under finite $D(2,1;a)$ transformations, its fails to produce a reasonable transition matrix $\mathcal{M}$ (see (\ref{CL})).

\vspace{0.5cm}

\noindent
{\bf 8. Conclusion}\\

\noindent
To summarise, in this work we extended a recent group--theoretic analysis of the $\mathcal{N}=4$ super--Schwarzian derivative \cite{GK}  to the case of the exceptional supergroup $D(2,1;a)$. The latter describes
the most general $\mathcal{N}=4$ supersymmetric extension of the conformal group in one dimension $SL(2,R)$. An analogue of the $\mathcal{N}=4$ super--Schwarzian derivative was built in terms of the Maurer--Cartan invariants. It was demonstrated that it lacked the conventional composition law unless extra nonlinear constraints were imposed upon the argument. The explicit solution of the constrains then showed that the natural candidate proved trivial and had to be superseded by one of the nonlinear constraints. An alternative candidate for a $D(2,1;a)$ super--Schwarzian derivative was proposed and its properties were established.
A parallel with the $\mathcal{N}=3$ case was drawn and a generalised $\mathcal{N}=3$ super--Schwarzian derivative was proposed.  A new $OSp(3|2)$ invariant was constructed.

Turning to possible further developments, it would be interesting to explore whether the alternative candidate for a $D(2,1;a)$ super--Schwarzian derivative proposed in Sect. 6 can be supported by superconformal field theory computations based upon an infinite--dimensional extension of $D(2,1;a)$ \cite{STVP}. A possibility to use it for constructing a $D(2,1;a)$ supersymmetric extension of the Sachdev--Ye--Kitaev model is worth studying as well.

As is known, the supergroups $SU(1,1|n)$, $Osp(n|2)$, and $Osp(4^{*}|2n)$ involve $SL(2,R)$ subgroup. It would be interesting to investigate whether consistent super--Schwarzian derivatives can be associated to them.
Note that a direct solution of superfield constraints may turn out to be problematic in that context, as the component decomposition becomes more involved with $n$ growing.

Another interesting open problem is a formulation of super--Schwarzian derivatives in superspace with universal cosmological attraction or repulsion \cite{FM}.

\vspace{0.5cm}

\noindent{\bf Acknowledgements}\\

\noindent
We thank S. Krivonos for reading an earlier version of the manuscript and useful comments.
This work is supported by the Russian Science Foundation, grant No 19-11-00005.

\vspace{0.5cm}

\noindent
{\bf Appendix A: Spinor conventions }

\vspace{0.5cm}

\noindent
Throughout the text we use a lower Greek index to designate an $SU(2)$--doublet representation. Hermitian conjugation
yields an equivalent representation to which one assigns an upper index
\be
{(\p_\a)}^{*}={\bar\p}^\a\ , \qquad \a=1,2\ .
\nonumber
\ee
As usual, spinor indices (including those on the covariant derivatives) are raised and lowered with the use of the $SU(2)$--invariant
Levi--Civita symbols
\be
\p^\a=\e^{\a\b}\p_\b\ , \quad {\bar\p}_\a=\e_{\a\b} {\bar\p}^\b\ ,
\nonumber
\ee
where $\e_{12}=1$, $\e^{12}=-1$. For spinor bilinears we stick to the notation
\be
\quad \p^2=(\p^\a \p_\a\ ) , \quad
\bar\p^2=(\bar\p_\a \bar\p^\a )\ , \quad \bar\p \p=(\bar\p^\a \p_\a )\ ,
\nonumber
\ee
such that
\bea
&&
\p_\a \p_\b=\frac 12 \e_{\a\b} \p^2\ , \qquad \bar\p^\a \bar\p^\b=\frac 12 \e^{\a\b} \bar\p^2\ , \qquad \p_\a \bar\p_\b-\p_\b \bar\p_\a=\e_{\a\b}
(\bar\p \p)\ ,
\nonumber\\[2pt]
&&
\psi^\alpha \psi^\beta=-\frac 12 \epsilon^{\alpha\beta} \psi^2, \qquad \bar\psi_\alpha \bar\psi_\beta=-\frac 12 \epsilon_{\alpha\beta} \bar\psi^2, \qquad {\left(\bar\psi \psi \right)}^2=\frac 12 \psi^2 \bar\psi^2.
\nonumber
\eea

The Pauli matrices ${{(\s_a)}_\a}^\b$
are taken in the standard form
\be
\s_1=\begin{pmatrix}0 & 1\\
1 & 0
\end{pmatrix}\ , \qquad \s_2=\begin{pmatrix}0 & -i\\
i & 0
\end{pmatrix}\ ,\qquad
\s_3=\begin{pmatrix}1 & 0\\
0 & -1
\end{pmatrix}\ ,
\nonumber
\ee
which obey
\begin{align}
&
{{(\s_a \s_b)}_\a}^\b +{{(\s_b \s_a)}_\a}^\b=2 \d_{ab} {\d_\a}^\b \ , &&
{{(\s_a \s_b)}_\a}^\b -{{(\s_b \s_a)}_\a}^\b=2i \e_{abc} {{(\s_c)}_\a}^\b \ ,
\nonumber\\[2pt]
&
{{(\s_a \s_b)}_\a}^\b=\d_{ab} {\d_\a}^\b +i \e_{abc} {{(\s_c)}_\a}^\b \ , &&
{{(\s_a)}_\a}^\b {{(\s_a)}_\g}^\r=2 {\d_\a}^\r {\d_\g}^\b-{\d_\a}^\b {\d_\g}^\r\ ,
\nonumber\\[2pt]
&
{{(\s_a)}_\a}^\b \e_{\b\g} ={{(\s_a)}_\g}^\b \e_{\b\a}\ , && \e^{\a\b} {{(\s_a)}_\b}^\g=\e^{\g\b} {{(\s_a)}_\b}^\a \ ,
\nonumber
\end{align}
where $\e_{abc}$ is the totally antisymmetric Levi-Civita symbol, $\e_{123}=1$. Throughout the text we use the abbriviation
$\bar\p \s_a \p=\bar\p^\a {{(\s_a)}_\a}^\b \p_\b$.

Our convention for the complex conjugation adopted above imply
\bea
&&
{(\bar\psi_\alpha)}^{*}=-\psi^\alpha\ , \qquad
{(\psi^2)}^{*}=\bar\psi^2\ , \qquad {(\bar\psi\,\sigma_a \chi)}^{*}=\bar\chi \sigma_a \psi\ , \qquad {\left(\epsilon_{\alpha\beta}\right)}^{*}=\epsilon^{\beta\alpha}.
\nonumber
\eea

\vspace{0.5cm}

\noindent
{\bf Appendix B: Lie superalgebra associated with $D(2,1;a)$}

\vspace{0.5cm}
\noindent
The structure relations of Lie superalgebra associated with the exceptional superconformal group $D(2,1;a)$ read
\begin{align}
&
[P,D] = iP, && [P,K] = 2iD,
\nonumber\\[2pt]
&
[D,K] = iK, && [\mathcal{J}_{b},\mathcal{J}_{c}] = i\epsilon_{bcd}\mathcal{J}_{d},
\nonumber\\[2pt]
&
\{Q_{\alpha},\bar{Q}^{\beta}\} = 2P{\delta_{\alpha}}^{\beta}, && \{Q_{\alpha},\bar{S}^{\beta}\} = -2i a {(\sigma_{c})_{\alpha}}^{\beta}\mathcal{J}_{c} - 2D{\delta_{\alpha}}^{\beta} + 2i(1+a)I_{3}{\delta_{\alpha}}^{\beta},
\nonumber\\[2pt]
&
\{S_{\alpha},\bar{S}^{\beta}\} = 2K{\delta_{\alpha}}^{\beta}, && \{\bar{Q}^{\alpha}, S_{\beta}\} = 2i a {(\sigma_{c})_{\beta}}^{\alpha}\mathcal{J}_{c} - 2D{\delta_{\beta}}^{\alpha} - 2i(1+a)I_{3}{\delta_{\beta}}^{\alpha},
\nonumber\\[2pt]
&
\{Q_{\alpha},S_{\beta}\} = -2(1+a)\epsilon_{\alpha\beta}I_{-}, && \{\bar{Q}^{\alpha},\bar{S}^{\beta}\} = 2(1+a)\epsilon^{\alpha\beta}I_{+},
\nonumber\\[2pt]
&
[D,Q_{\alpha}] = -\frac{i}{2}Q_{\alpha}, && [D,S_{\alpha}] = \frac{i}{2}S_{\alpha},
\nonumber\\[2pt]
&
[K,Q_{\alpha}] = iS_{\alpha}, && [P,S_{\alpha}] = -iQ_{\alpha},
\nonumber\\[2pt]
&
[\mathcal{J}_{c},Q_{\alpha}] = -\frac{1}{2}{(\sigma_{c})_{\alpha}}^{\beta}Q_{\beta}, && [\mathcal{J}_{c},S_{\alpha}] = -\frac{1}{2}{(\sigma_{c})_{\alpha}}^{\beta}S_{\beta},
\nonumber\\[2pt]
&
[D,\bar{Q}^{\alpha}] = -\frac{i}{2}\bar{Q}^{\alpha}, && [D,\bar{S}^{\alpha}] = \frac{i}{2}\bar{S}^{\alpha},
\nonumber\\[2pt]
&
[K,\bar{Q}^{\alpha}] = i\bar{S}^{\alpha}, && [P,\bar{S}^{\alpha}] = -i\bar{Q}^{\alpha},
\nonumber\\[2pt]
&
[\mathcal{J}_{c},\bar{Q}^{\alpha}] = \frac{1}{2}\bar{Q}^{\beta}{(\sigma_{c})_{\beta}}^{\alpha}, && [\mathcal{J}_{c},\bar{S}^{\alpha}] = \frac{1}{2}\bar{S}^{\beta}{(\sigma_{c})_{\beta}}^{\alpha},
\nonumber\\[2pt]
&
[I_{-},\bar{Q}^{\alpha}] = i\epsilon^{\alpha\beta}Q_{\beta}, && [I_{-},\bar{S}^{\alpha}] = i\epsilon^{\alpha\beta}S_{\beta},
\nonumber
\end{align}
\begin{align}
&
[I_{+},Q_{\alpha}] = -i\epsilon_{\alpha\beta}\bar{Q}^{\beta}, && [I_{+},S_{\alpha}] = -i\epsilon_{\alpha\beta}\bar{S}^{\beta},
\nonumber\\[2pt]
&
[I_{3},Q_{\alpha}] = -\frac{1}{2}Q_{\alpha}, && [I_{3},S_{\alpha}] = -\frac{1}{2}S_{\alpha},
\nonumber\\[2pt]
&
[I_{3},\bar{Q}^{\alpha}] = \frac{1}{2}\bar{Q}^{\alpha}, && [I_{3},\bar{S}^{\alpha}] = \frac{1}{2}\bar{S}^{\alpha},
\nonumber\\[2pt]
&
[I_{-},I_{3}] = I_{-}, && [I_{+},I_{3}] = -I_{+},
\nonumber\\[2pt]
&
[I_{-},I_{+}] = -2I_{3}.
\nonumber
\end{align}
Here $a$ is a real parameter and ${{(\s_c)}_\b}^\a$ are the Pauli matrices. Note that for $a=-1$ the superalgebra reduces to $su(1,1|2)\oplus su(2)$.

\vspace{0.5cm}

\noindent
{\bf Appendix C: Maurer--Cartan invariants}

\vspace{0.5cm}
\noindent
In this Appendix, we display the Maurer--Cartan invariants which result from the group--theoretic element (\ref{gte}) and the prescription (\ref{MC})
\bea
&&
{\left(\omega_{P}\right)}^{\alpha}= e^{-\nu}\left(\mathcal{D}^{\alpha}\rho - i\psi_{\beta}\mathcal{D}^{\alpha}\bar{\psi}^{\beta} -i \bar{\psi}^{\beta}\mathcal{D}^{\alpha}\psi_{\beta}\right),
\nonumber
\\[2pt]
&&
{\left(\omega_{D}\right)}^{\alpha} = \mathcal{D}^{\alpha}\nu+ 2i\left(\phi_{\beta}\mathcal{D}^{\alpha}\bar{\psi}^{\beta} + \bar{\phi}^{\beta}\mathcal{D}^{\alpha}\psi_{\beta}\right)
-2\mu e^{\nu} {\left(\omega_{P}\right)}^{\alpha},
\nonumber
\\[2pt]
&&
{\left(\omega_{K}\right)}^{\alpha} = e^{\nu} \left[\mathcal{D}^{\alpha}\mu - i \left(\phi_{\beta}\mathcal{D}^{\alpha}\bar{\phi}^{\beta} + \bar{\phi}^{\beta}\mathcal{D}^{\alpha}\phi_{\beta}\right) - 2i\mu  \left(\phi_{\beta}\mathcal{D}^{\alpha}\bar{\psi}^{\beta} + \bar{\phi}^{\beta}\mathcal{D}^{\alpha}\psi_{\beta}\right)
\right.
\nonumber
\\[2pt]
&&
\left.
\qquad \qquad
+2(1+2 a) \bar{\phi}\phi \left(\bar{\phi}^{\beta}\mathcal{D}^{\alpha}\psi_{\beta} - \phi_{\beta}\mathcal{D}^{\alpha}\bar{\psi}^{\beta}\right)\right]+ e^{2\nu} \left(\mu^{2} - \frac{(1+2 a)}{2}\phi^{2}\bar{\phi}^{2}\right) {\left(\omega_{P}\right)}^{\alpha},
\nonumber
\\[2pt]
&&
{\left(\omega_{Q}\right)}^{\alpha\beta} = e^{\frac{ik_{3}-\nu}{2}}\left[\left(\mathcal{D}^{\alpha}\psi^{\gamma}\right)\cos{k} - \left(\mathcal{D}^{\alpha}\bar{\psi}^{\gamma}\right)\frac{k\sin{k}}{k_{+}} + e^{\nu} \left(\phi^{\gamma}\cos{k} - \bar{\phi}^{\gamma}\frac{k\sin{k}}{k_{+}}\right){\left(\omega_{P}\right)}^{\alpha} \right]{R_{\gamma}}^{\beta},
\nonumber
\\[2pt]
&&
{\left(\omega_{\bar{Q}} \right)}_\beta{}^{\alpha} = e^{-\frac{ik_{3} + \nu}{2}}\left[\left(\mathcal{D}^{\alpha}\bar{\psi}_{\gamma}\right)  \cos{k} + \left(\mathcal{D}^{\alpha}\psi_{\gamma}\right) \frac{k\sin{k}}{k_{-}}
+ e^{\nu} \left(\bar{\phi}_{\gamma}\cos{k} + \phi_{\gamma}\frac{k\sin{k}}{k_{-}}\right) {\left(\omega_{P}\right)}^{\alpha}
\right]{{\bar{R}_{\beta}}}^{\phantom{\beta}\gamma},
\nonumber
\\[2pt]
&&
{\left(\omega_{S}\right)}^{\alpha\beta} = \mu e^{\nu}{\left(\omega_{Q}\right)}^{\alpha\beta} - e^{\frac{ik_{3}+\nu}{2}}\left[i \bar{\phi}\phi\left( \left(\mathcal{D}^{\alpha}\psi^{\gamma} \right)\cos{k} + \left(\mathcal{D}^{\alpha}\bar{\psi}^{\gamma}\right)\frac{k\sin{k}}{k_{+}}\right)
\right.
\nonumber
\\[2pt]
&&
\qquad \qquad
+ 2i(1+a)\left(\phi_{\xi} \left(\mathcal{D}^{\alpha}\psi^{\xi} \right) \bar{\phi}^{\gamma}\cos{k} + \bar{\phi}_{\xi} \left(\mathcal{D}^{\alpha}\bar{\psi}^{\xi}\right) \phi^{\gamma}\frac{k\sin{k}}{k_{+}}\right)
\nonumber
\\[2pt]
&&
\qquad \qquad
- 2ia\left(\phi_{\xi} \left(\mathcal{D}^{\alpha}\bar{\psi}^{\xi} \right) \phi^{\gamma}\cos{k} - \bar{\phi}^{\xi} \left( \mathcal{D}^{\alpha}\psi_{\xi} \right) \bar{\phi}^{\gamma}\frac{k\sin{k}}{k_{+}}\right)
\nonumber
\\[2pt]
&&
\qquad \qquad
\left.
- \left(\left(\mathcal{D}^{\alpha}\phi^{\gamma} \right) \cos{k} - \left(\mathcal{D}^{\alpha}\bar{\phi}^{\gamma} \right) \frac{k\sin{k}}{k_{+}}\right)\right] {R_{\gamma}}^{\beta}
\nonumber
\\[2pt]
&&
\qquad \qquad
+ i(1+2 a) e^{\frac{ik_{3}+3\nu}{2}}\bar{\phi}\phi\left(\phi^{\gamma}\cos{k} + \bar{\phi}^{\gamma}\frac{k\sin{k}}{k_{+}}\right) {R_{\gamma}}^{\beta} {\left(\omega_{P}\right)}^{\alpha},
\nonumber\\[2pt]
&&
{\left(\omega_{\bar{S}} \right)}_\beta{}^{\alpha} = \mu e^{\nu}{\left(\omega_{\bar{Q}}\right)}_\beta{}^{\alpha} + e^{\frac{-ik_{3} + \nu}{2}} \left[i\bar{\phi}\phi\left( \left(\mathcal{D}^{\alpha}\bar{\psi}_{\gamma}\right)\cos{k} - \left(\mathcal{D}^{\alpha}\psi_{\gamma}\right) \frac{k\sin{k}}{k_{-}}\right)
\right.
\nonumber
\eea
\bea
&&
\qquad \qquad
- 2i(1+a) \left(\bar{\phi}^{\xi} \left(\mathcal{D}^{\alpha}\bar{\psi}_{\xi} \right) \phi_{\gamma} \cos{k} - \phi^{\xi} \left(\mathcal{D}^{\alpha}\psi_{\xi} \right) \bar{\phi}_{\gamma} \frac{k\sin{k}}{k_{-}}\right)
\nonumber\\[2pt]
&&
\qquad \qquad
+ 2ia \left(\bar{\phi}^{\xi} \left(\mathcal{D}^{\alpha}\psi_{\xi} \right) \bar{\phi}_{\gamma} \cos{k} + \phi_{\xi} \left(\mathcal{D}^{\alpha}\bar{\psi}^{\xi} \right) \phi_{\gamma}\frac{k\sin{k}}{k_{-}}\right)
\nonumber
\\[2pt]
&&
\qquad \qquad
\left.
+\left(\left(\mathcal{D}^{\alpha}\bar{\phi}_{\gamma}\right)\cos{k} + \left(\mathcal{D}^{\alpha}\phi_{\gamma}\right)\frac{k\sin{k}}{k_{-}}\right)\right]{\bar{R}_{\beta}}^{\phantom{\beta}\gamma}
\nonumber\\[2pt]
&&
\qquad \qquad
-i(1+2a) e^{\frac{-ik_{3} + 3\nu}{2}}\bar{\phi}\phi\left(\bar{\phi}_{\gamma}\cos{k} - \phi_{\gamma}\frac{k\sin{k}}{k_{-}}\right){\bar{R}_{\beta}}^{\phantom{\beta}\gamma} {\left(\omega_{P}\right)}^{\alpha},
\nonumber\\[2pt]
&&
{\left(\omega_{\mathcal{J}}\right)}_c{}^{\alpha} = \left(\mathcal{D}^{\alpha}\lambda_{c}\right)\frac{\sin{\lambda}}{\lambda} - \lambda_{c}\lambda_{b} \left(\mathcal{D}^{\alpha}\lambda_{b}\right)\frac{\sin{\lambda} - \lambda}{\lambda^{3}} + \epsilon_{cbd}\lambda_{b} \left(\mathcal{D}^{\alpha}\lambda_{d}\right)\frac{1-\cos{\lambda}}{\lambda^{2}}
\nonumber
\\[2pt]
&&
\qquad \qquad
+ 2a{(\sigma_{b})_{\gamma}}^{\xi} \left(e^{\nu}\phi^{\gamma}\bar{\phi}_{\xi}{\left(\omega_{P}\right)}^{\alpha}+ \left(\mathcal{D}^{\alpha}\psi^{\gamma}\right)\bar{\phi}_{\xi} - \left(\mathcal{D}^{\alpha}\bar{\psi}_{\xi}\right)\phi^{\gamma}\right) \times
\nonumber
\\[2pt]
&&
\qquad \qquad
\times
\left(\delta_{cb}\cos{\lambda} - \epsilon_{cbd}\lambda_{d}\frac{\sin{\lambda}}{\lambda} + \lambda_{c}\lambda_{b}\frac{1-\cos{\lambda}}{\lambda^{2}}\right),
\nonumber
\\[2pt]
&&
{\left(\omega_{-}\right)}^{\alpha} = e^{ik_{3}}\left[\mathcal{D}^{\alpha}k_{-} - \frac{2k^2 - k\sin{2k}}{4k_{+}}\left(\frac{\mathcal{D}^{\alpha}k_{-}}{k_{-}} - \frac{\mathcal{D}^{\alpha}k_{+}}{k_{+}}\right)
\right.
\nonumber
\\[2pt]
&&
\qquad \qquad
\left.
+ i (1+a)\left( \frac{\left(\mathcal{D}^{\alpha}\bar{\psi}_{\beta}\right)}{k_{+}} \left(2 \bar{\phi}^{\beta} k_{-}\sin^{2}{k} - \phi^{\beta}k\sin{2k}\right)
\right.
\right.
\nonumber
\\[2pt]
&&
\qquad \qquad
\left.
\left.
-\left(\mathcal{D}^{\alpha}\psi^{\beta}\right)\left(2\phi_{\beta}\cos^{2}{k} - \bar{\phi}_{\beta}\frac{k\sin{2k}}{k_{+}}\right)\right)\right]
\nonumber
\\[2pt]
&&
\qquad \qquad
- i (1+a)e^{ik_{3} + \nu} \left(\phi^{2}\cos^{2}{k} - \bar{\phi}^{2}\frac{k_{-}\sin^{2}{k}}{2k_{+}} - \bar{\phi}\phi\frac{k\sin{2k}}{k_{+}}\right) {\left(\omega_{P}\right)}^{\alpha},
\nonumber\\[2pt]
&&
{\left(\omega_{+}\right)}^{\alpha} = e^{-ik_{3}}\left[\mathcal{D}^{\alpha}k_{+} - \frac{2k^{2} - k\sin{2k}}{4k_{-}}\left(\frac{\mathcal{D}^{\alpha}k_{+}}{k_{+}} - \frac{\mathcal{D}^{\alpha}k_{-}}{k_{-}}\right)
\right.
\nonumber
\\[2pt]
&&
\qquad \qquad
\left.
- i (1+a)\left(\frac{\left(\mathcal{D}^{\alpha}\psi^{\beta}\right)}{k_{-}}\left(2\phi_{\beta}k_{+}\sin^{2}{k} + \bar{\phi}_{\beta} k\sin{2k}\right)
\right.
\right.
\nonumber
\\[2pt]
&&
\qquad \qquad
\left.
\left.
+ \left(\mathcal{D}^{\alpha}\bar{\psi}^{\beta}\right)\left(2\bar{\phi}_{\beta}\cos^{2}{k} + \phi_{\beta}\frac{k\sin{2k}}{k_{-}}\right)\right)\right]
\nonumber
\\[2pt]
&&
\qquad \qquad
+ i (1+a)e^{-ik_{3} + \nu}\left(\bar{\phi}^{2}\cos^{2}{k} - \phi^{2}\frac{k_{+}\sin^{2}{k}}{k_{-}} - \bar{\phi}\phi\frac{k\sin{2k}}{k_{-}}\right){\left(\omega_{P}\right)}^{\alpha},
\nonumber
\\[2pt]
&&
{\left(\omega_{3}\right)}^{\alpha} = \mathcal{D}^{\alpha}k_{3} + \frac{i\sin^{2}{k}}{k^{2}}\left(k_{-}\mathcal{D}^{\alpha}k_{+} -  k_{+}\mathcal{D}^{\alpha}k_{-}\right)
+ 2(1+a)\left(\mathcal{D}^{\alpha}\psi_{\beta}\right)\left(\bar{\phi}^{\beta} \cos{2k} + \phi^{\beta} \frac{k\sin{2k}}{k_{-}}\right)
\nonumber\\[2pt]
&&
\qquad \qquad
 - 2(1+a)\left(\mathcal{D}^{\alpha}\bar{\psi}^{\beta}\right)\left(\phi_{\beta} \cos{2k} - \bar{\phi}_{\beta} \frac{k\sin{2k}}{k_{+}}\right)
 \nonumber
\\[2pt]
&&
\qquad \qquad
 -(1+a)e^{\nu}\sin{2k}\left(\frac{k\phi^{2}}{k_{-}} + \frac{k\bar{\phi}^{2}}{k_{+}} + 2\bar{\phi}\phi\cot{2k}\right){\left(\omega_{P}\right)}^{\alpha},
\nonumber
\eea
where
$\lambda = \sqrt{\lambda_{a}\lambda_{a}}$, $k = \sqrt{k_{-}k_{+}}$, and ${R_{\alpha}}^{\beta} = {{\left(\mbox{exp} \left[\frac{i}{2} \lambda_a \sigma_a \right] \right)}_\alpha}^\beta$,
${{\bar R}_{\alpha}}{}^{\beta} = {\left(\exp{\left[-\frac{i}{2}\lambda_{c}\sigma_{c}\right]}\right)_{\alpha}}^{\beta}$.

\end{document}